\documentclass[fleqn,usenatbib]{mnras}

\usepackage{newtxtext,newtxmath}

\usepackage[T1]{fontenc}

\DeclareRobustCommand{\VAN}[3]{#2}
\let\VANthebibliography\thebibliography
\def\thebibliography{\DeclareRobustCommand{\VAN}[3]{##3}\VANthebibliography}

\usepackage{graphicx}	
\usepackage{amsmath}	

\usepackage[normalem]{ulem} 
\usepackage[dvipsnames]{xcolor} 
\usepackage{wasysym} 

\newcommand{\eg}[0]{$\textnormal{e.g. }$}
\newcommand{\ie}[0]{$\textnormal{i.e. }$}
\newcommand{\tn}[1]{\textnormal{#1}}
\newcommand{\sub}[1]{_{\textnormal{#1}}}
\newcommand{\Msun}[0]{\,\textnormal{M}_{\textnormal{\astrosun}}}

\title[Stellar migration in Auriga]{Stellar migration in the Auriga simulations}

\author[P. Okalidis et al.]{Periklis Okalidis,$^{1}$\thanks{okalidis@mpa-garching.mpg.de}
Robert J. J. Grand,$^{1,2,3}$
Robert M. Yates,$^{4}$
Volker Springel$^{1}$
\\
$^{1}$Max-Planck-Institut f{\"u}r Astrophysik, D-85741 Garching, Germany\\
$^2$Instituto de Astrof\'isica de Canarias, Calle Vía L\'actea s/n, E-38205 La Laguna, Tenerife, Spain\\
$^3$Departamento de Astrof\'isica, Universidad de La Laguna, Av. del Astrof\'isico Francisco S\'anchez s/n, E-38206, La Laguna, Tenerife, Spain\\
$^{4}$Department of Physics, University of Surrey, Stag Hill, Guildford, GU2 7XH, UK
}

\date{Accepted XXX. Received YYY; in original form ZZZ}

\pubyear{2022}

\begin{document}
\label{firstpage}
\pagerange{\pageref{firstpage}--\pageref{lastpage}}
\maketitle

\begin{abstract}
We study the presence and importance of stellar migration in the evolution of 17 Milky-Way like disk galaxies with stellar mass $10 < \textrm{log}(M_{*}/{\rm M}_\odot) < 11$ from the Auriga suite of zoom-in cosmological hydrodynamical simulations. We compare the birth radii of the stars to their radii at $z=0$ for each system and present mean values of the strength of stellar migration as a function of radius and stellar age which vary between 1-4 kpc. We also investigate the effect of migration on age and metallicity radial profiles in the disks. We find several cases of age gradient flattening due to migration, but significant changes to metallicity profiles only for older stellar populations and disks that develop a strong bar. Furthermore, we study stellar migration from the perspective of the change of the galactocentric radius ($\Delta R$) and orbital guiding centre radius ($\Delta R_g$) of stellar particles between given time intervals.  We find that stars migrate approximately as a diffusion process only in the outer parts of the disks and for particular galaxies that have a weak bar. Strongly barred galaxies in our sample show larger stellar migration but its timestep evolution is slower-than-diffusion. Finally, we give parametrisations that encapsulate the dependence of the strength of the radial migration as a function of time and radius, for incorporation into (semi-)analytic models of galaxy evolution.
\end{abstract}




\color{black}
\section{Introduction}

During the lifetime of a star its orbital radius within the galactic plane is subject to changes that can result in the star inhabiting a radius different than the one it was born at, a concept referred as stellar radial migration \citep[e.g.][]{LyndenBell72}. There are two terms that are widely discussed in the literature, describing two entirely distinct types of stellar migration, `churning' and `blurring' \citep{Schoenrich09}.`Churning' refers to the direct change of the guiding centre, the mean radius of the stellar orbit, and relates to permanent changes in the orbital angular momentum without changing the ``random'' component of its orbital energy \citep[e.g.][]{Grand12}. `Blurring' is associated with temporary changes in the orbital kinetic energy close to peri/apo-centre (thus away from the mean orbital radius), while the angular momentum remains constant. There are several mechanisms that are responsible for inducing these changes in the orbital radii, including non-axisymmetric features in the galactic disk, such as a bar \citep[e.g.][]{Halle18}, transient spiral arms \citep[e.g.][]{Sellwood02} and interactions with giant molecular clouds (GMCs). Furthermore, minor mergers with satellite systems have also been explored as drivers of radial migration \citep[e.g.][]{Quillen09}. However, disentangling all of these mechanisms is a far from trivial task.

Radial migration has been invoked in order to potentially explain many observables in the Milky Way, such as planar dynamical streams \citep{Kawata2018,Hunt2018}, the large spread of stellar metallicities \citep[e.g.][]{Nordstrom14,Haywood2008,Kubryk13,Minchev13,Grand15} and the presence of supersolar metallicity stars \citep{Kordopatis15} in the solar neighbourhood as well as the large scatter in the age-metallicity relation \citep{Casagrande11}, although recent studies \citep{Haywood2013,Bergemann14,Walcher16} point to an age-metallicity anti-correlation especially for stars older than $\sim{}9$ Gyr. Additionally, migration is a possible mechanism that may explain disk truncations and the upturn of age gradients observed in the outer regions of galaxies \citep[e.g.][]{Bakos2008,Roskar08b,Radburn12,RuizLara2017,Herpich2017} and the bi-modality in the  [$\alpha$/Fe]-[Fe/H] relation in the Milky Way, which has been studied both from an observational \citep{Fuhrmann98,Haywood2013,Anders14,Nidever14,Hayden15} and theoretical \citep{Schoenrich09,Brook12,Minchev14,Grand18,Mackereth19,Clarke19,Buck20,Renaud21,Khoperskov21} perspective. Radial migration has been linked also to the formation of the geometrically-defined galactic thick disk \citep{Loebman11,Solway12,VeraCiro14} and shaping vertically ``flared'' distributions of coeval stellar populations \citep[e.g.][]{Minchev15}. However, there is also contrasting evidence that stellar migration is ineffective in thick disk formation and rather contributes to cooling the disk in the cosmological context \citep{Minchev12b,Minchev14,Grand16,Ma17}. Observational studies of the migration process are non trivial, since it is not possible to get direct information about the initial (birth) conditions of a single star's orbit. Rather, migration can be inferred indirectly from measuring metallicity and age gradients and identifying different regions in the age-metallicity plane. One useful tracer of stellar migration in the Milky Way are stellar clusters.  \cite{Netopil21} used the ages and metallicities of a number of open clusters to measure metallicity gradients and inferred a mean migration rate of $1\, {\rm kpc/Gyr}^{-1}$ for younger objects and half of this value for older objects.

Radial migration has been consistently studied in chemodynamical models of disk galaxies that aim to reproduce the observed age-metallicity relations and the radial gradients of these quantities in the solar neighbourhood. In models such as \cite{Sellwood02} and \cite{Minchev14}, radial migration is treated using self-consistent angular momentum redistribution from an N-body disk, whereas \cite{Schoenrich09} and \cite{Kubryk13,Kubryk15} add prescriptions that describe separately the churning and blurring processes. In their chemodynamical model, \cite{Frankel18} introduce an analytic formulation for stellar migration that follows a Gaussian diffusion process where older stars spread to increasingly larger radii from their birth radius. In their formulation, the overall shape of the Gaussian function is regulated by a single migration strength parameter. APOGEE data was used to fit the best parameters for their model \citep{Majewski17}. In a later study \citep{Frankel20}, their model was expanded to study the effect of churning and blurring separately, concluding that churning has an order of magnitude stronger effect than blurring. \cite{Johnson21} use data from a numerical simulation and test different prescriptions for their chemodynamical model to study the migration process, in a similar way to \cite{Minchev13}.

Similarly, several studies have been carried out using numerical simulations to study the presence and strength of the migration process in simulated disk galaxies. These studies can be split between those of relying on isolated disk systems (\eg{}\citealt{DiMatteo13,Aumer16, Halle18, Mikkola20}) and those that analyse disks embedded in a cosmological environment using zoom-in simulations (\eg{}\citealt{Martig14,Grand16,Buck20}). The advantage of the former is that they can isolate and study in a controlled setup the effect of non-axisymmetries forming during the lifetime of the disk have on the stellar orbits. The latter, though, create a more realistic analogue of a real galaxy where the deviation of the positions of stars from their birth radii is the cumulative result of not only all the angular momentum changes induced by bars or spiral arms but also by any merger events that may have happened to the particular system. 

\cite{Minchev10} explored the combined effect arising from overlapping resonances when both a bar and spiral arms are present in a simulated disk. \cite{Minchev12a} concluded from their numerical simulations that the effect of bars is dominant with regards to stellar migration compared to the effect of transient spiral arms, and also confirm the importance of migration for the flattening and reversal of age gradients in the disk outskirts which are found to be populated by stars that have transferred outwards from the inner parts of the disk. Similarly, \cite{Agertz21} confirmed that metal rich stars preferentially migrate from inner to outer regions in their simulated Milky Way galaxy, affecting the metallicity distribution function around the solar neighbourhood. Finally, \cite{Verma21} applied sophisticated forward-modelling techniques to halos from the Auriga cosmological simulations to put constraints on the strength of migration based on measures of the metallicity dispersion at the Solar cylinder.

In this project, we study the process of radial migration of stars in a number of different Milky Way-mass halos from the high-resolution cosmological zoom-in simulation suite Auriga \citep{GGM17}, which includes environmental effects such as mergers and gas accretion. We study the total migration of the stars over their lifetime, measure the migration strength in each galaxy model, and also compare the simulated profiles of age and metallicity with fictitious profiles that would result if there was no evolution in the positions of the stellar particles. In a separate analysis we look at the migration of stars between different output snapshots of the simulations, and we arrive at a simple parameterisation for the stellar radial migration in Auriga that can be easily incorporated into (semi-)analytic models of galaxy evolution.

This paper is structured as follows. In Section~2, we briefly review the simulation suite we use for this study. In Section~3 we analyze the strength of migration from the perspective of the birth radius of stars, while in Section~4 we focus on the rate of migration by comparing stellar positions in subsequent simulation outputs.
Finally, we give a discussion or our results and summarize our conclusions in Section~5.

\section{Simulations}

We make use of the Auriga suite of high-resolution, magneto-hydrodynamical cosmological ``zoom-in'' simulations \citep{GGM17} which are designed to reproduce Milky-Way analogue disk galaxies in the concordance $\Lambda$CDM cosmology.  We select a total of 17 Auriga halos; 9 halos from the original runs of the project with a halo mass\footnote{Defined to be the mass inside a sphere in which the mean matter density is 200 times the critical density of the Universe, $\rho _{\rm crit} = 3H^2(z)/(8 \pi G)$.} ranging between $1-2 \times 10^{12}\, {\rm M}_{\odot}$, and 8 halos of a recent lower mass extension in the range $0.5-1 \times 10^{12}\, {\rm M}_{\odot}$ \citep{GVZ19}. 

These halos were originally selected based on a mild isolation criterion in the $z=0$ snapshot of the dark matter-only counterpart to the cosmological EAGLE simulation of co-moving side length $67.8\,h^{-1} {\rm cMpc}$  (L100N1504) presented in \citet{SCB15}. The values of the cosmological parameters for these simulations are $\Omega\sub{m}=0.307$, $\Omega\sub{b}=0.048$, $\Omega_{\Lambda}=0.693$, $H_{0}=100\,h\, {\rm km}\, {\rm s}^{-1}\, {\rm Mpc}^{-1}$ and $h=0.667$, taken from \citet{PC13}.

The zoom simulations are initialized at redshift $z=127$ with the high-resolution regions having a mass resolution of $\sim 5\times10^{4}$ $\rm M_{\odot}$ per baryonic element and a comoving softening length of $500\, h^{-1} {\rm pc} $. The physical softening length grows until $z=1$, after which time it is kept fixed. The physical softening value for the gas cells is scaled by the gas cell radius (assuming a spherical cell shape given the volume), with a minimum softening set to that of the collisionless particles.

Unlike in EAGLE, the time evolution in Auriga is carried out with the quasi-Lagrangian magneto-hydrodynamics simulation code {\small AREPO} \citep{Sp10,PSB15,Weinberger20}, using the galaxy formation model outlined in \citet{GGM17} that accounts for the most important physical processes  relevant for galaxy formation and evolution. In {\small AREPO}, gas cells are modelled with an unstructured mesh in which gas cells move with the local bulk flow. This numerical approach combines the accuracy of a mesh-based representation of hydrodynamics with the geometrical flexibility and low advection errors of a Lagrangian treatment. 

\subsection{Included galaxy formation physics}

The physical processes incorporated into Auriga's galaxy formation model include primordial and metal-line cooling  \citep{VGS13}, as well as an externally imposed spatially uniform UV  background for modelling cosmic reionization. The star-forming interstellar medium (ISM) comprises gas that has become denser than $0.11$ atoms $\rm cm^{-3}$ and is modelled by a subgrid model that describes a two phase medium of cold clouds embedded in a hot volume filling phase \citep{SH03} assumed to be in pressure equilibrium. 

Stellar particles are spawned stochastically from the gas using a Schmidt-type star formation prescription with a gas consumption timescale calibrated to observed star formation densities. Each particle represents a Simple Stellar Population (SSP) characterised by properties such as its age, mass and metallicity. The stellar evolution model applied to each SSP follows type Ia supernovae (SNe-Ia) and winds from Asymptotic Giant Branch (AGB) stars that return mass and metals (9 elements are tracked: H, He, C, O, N, Ne, Mg, Si and Fe) to the surrounding gas. Supernovae type II (SNe-II) are also assumed to return mass and metals, but are treated with an instantaneous recycling approximation.

Galactic winds from SNe-II are modelled by a wind particle scheme mediating non-local kinetic feedback \citep{VGS13}, which effectively models the removal of mass from star-forming regions and deposits mass, momentum and energy into adjacent gas in the circumgalactic medium with density lower than $5\%$ of the density of star-forming gas. These winds are an important feedback channel for regulating the total stellar mass forming in the galaxies.

In addition, there are prescriptions in the model accounting for the accretion of matter onto black holes and energetic feedback from Active Galactic Nuclei \citep[as described in][]{GGM17}. Also, magnetic fields are seeded at $z=127$ with a co-moving field strength of $10^{-14}$ cG \citep{PMS14}, and are subsequently amplified by small-scale dynamo processes during the simulations. While several studies have shown that the resulting magnetic field strength evolution and radial profile in Milky Way-like halos are in good agreement between Auriga and observational findings \citep[e.g][]{PGG17,PGP18,Pakmor20}, they play only a minor role for the regulation of star formation in Auriga, and thus are probably of negligible influence on the stellar migration rates. Similarly, the impact of central supermassive black holes is probably minor, at least in the outer part of disks, whereas their feedback  can indirectly have an impact in the inner regions through influencing the mass of the stellar bulge and bar properties \citep[see][]{Irodotou22}.

In our runs, we have 252 time-slice outputs (`snapshots') down to redshift $z=0$, with a median time resolution of $\sim 60\, {\rm Myr}$ between two consecutive snapshots (the time interval ranges between 45-75 Myr). Between the different simulated galaxies, we observe a variety of structural properties, with nearly half of them developing a bar at some point in their evolution. In addition, the disks have varying merger histories, with the lower-mass halos experiencing more significant merger events at lower redshift, whereas the higher-mass ones have no significant mergers in the last 3 Gyr of the simulation.

\subsection{Galactic properties}

\begin{table*}
\centering
\begin{tabular}{|c|c|c|c|c|c|c|c|c|c|}
\hline
  \multicolumn{1}{|c|}{Name} &
  \multicolumn{1}{c|}{log$(M_{*}/{\rm M}_{\odot})$} &
  \multicolumn{1}{c|}{$R_{50}\;$(kpc)} &
  \multicolumn{1}{c|}{$R_{90}\;$(kpc)} &
  \multicolumn{1}{c|}{A2$\sub{max}$} &
  \multicolumn{1}{c|}{$\langle t_{*,\tn{age}} \rangle \;$(Gyr) } &
  \multicolumn{1}{c|}{$v\sub{rot,max} $ (km s$^{-1}$)} &
  \multicolumn{1}{c|}{$\langle v_{*} \rangle \;$ (km s$^{-1}$)} &
  \multicolumn{1}{c|}{$\sigma_{*}$ (km s$^{-1}$)} &
  \multicolumn{1}{c|}{$\langle v_{*} \rangle / \sigma_{*}$} \\
\hline
  halo\_5 & 10.81 & 3.31 & 11.46 & 0.421 & 5.4 & 265.4 & 231.2 & 74.4 & 3.11\\
  halo\_6 & 10.63 & 6.42 & 14.65 & 0.18 & 6.3 & 209.7 & 187.2 & 55.5 & 3.37\\
  halo\_9 & 10.81 & 2.83 & 8.4 & 0.449 & 6.3 & 276.3 & 241.9 & 81.5 & 2.97\\
  halo\_13 & 10.66 & 3.11 & 7.8 & 0.208 & 4.4 & 245.2 & 218.5 & 72.3 & 3.02\\
  halo\_17 & 10.88 & 1.71 & 9.24 & 0.498 & 6.8 & 345.1 & 275.2 & 98.3 & 2.8\\
  halo\_23 & 10.87 & 6.52 & 15.02 & 0.198 & 6.3 & 256.7 & 229.8 & 71.2 & 3.23\\
  halo\_24 & 10.79 & 4.59 & 15.71 & 0.503 & 7.0 & 246.0 & 212.8 & 70.7 & 3.01\\
  halo\_26 & 10.95 & 4.21 & 12.42 & 0.501 & 5.9 & 288.5 & 256.6 & 82.1 & 3.13\\
  halo\_28 & 10.93 & 3.25 & 8.58 & 0.301 & 4.5 & 299.8 & 270.3 & 90.0 & 3.0\\
  halo\_L1 & 10.05 & 5.07 & 13.88 & 0.097 & 4.4 & 150.8 & 125.6 & 47.1 & 2.67\\
  halo\_L2 & 10.20 & 3.85 & 13.46 & 0.319 & 5.1 & 161.7 & 145.9 & 62.3 & 2.34\\
  halo\_L3 & 10.45 & 4.05 & 12.32 & 0.308 & 6.2 & 200.5 & 177.8 & 69.5 & 2.56\\
  halo\_L5 & 10.13 & 4.31 & 12.0 & 0.157 & 5.6 & 167.5 & 139.4 & 55.5 & 2.51\\
  halo\_L7 & 10.36 & 3.95 & 13.9 & 0.089 & 5.2 & 172.6 & 153.8 & 47.4 & 3.24\\
  halo\_L8 & 10.59 & 4.55 & 11.28 & 0.307 & 5.2 & 210.8 & 185.8 & 60.2 & 3.09\\
  halo\_L9 & 10.32 & 3.77 & 10.88 & 0.311 & 7.0 & 168.8 & 149.1 & 44.3 & 3.37\\
  halo\_L10 & 10.42 & 3.75 & 12.56 & 0.254 & 7.1 & 189.7 & 168.1 & 65.5 & 2.57\\
\hline\end{tabular}
\caption{Properties of the different galactic disks at $z=0$. We show in order the stellar mass enclosed within 2 kpc from the plane and 20 kpc in radius ($M_{*}$), characteristic radii of 50\% and 90\% of the stellar mass ($R_{50}, R_{90}$), the maximum A2 coefficient, mean stellar age ($\langle t_{*,\tn{age}} \rangle$), maxiumum rotational velocity ($v\sub{rot,max}$), mean stellar velocity ($\langle v_{*} \rangle$), stellar velocity dispersion ($\sigma_{*}$), and the ratio of the velocity to the velocity dispersion.}
\label{proptable}
\end{table*}

In Table \ref{proptable}, we summarize some of the $z=0$ properties of the 17 simulated galaxy disks which we consider relevant in this study. Halos belonging to the lower halo mass simulations are named with prefix 'L'. The stellar masses in the Table refer to the mass enclosed within 2 kpc of the disk plane and within 20 kpc in galactocentric radius and ranges between $10 < \textrm{log}(M_{*} / {\rm M}_{\odot}) < 11$.  We also report the radii which enclose 50 per cent ($R_{50}$) and 90 per cent ($R_{90}$) of the disk stellar mass. Because our disks have a variety of sizes, we use scaled radii in the presentation of many results in the next sections. We have tested three options for the scaling, (1) the disk scale length taken from the slope in a power law fit of the stellar density, (2) $R_{90}$ and (3) $R_{50}$. $R_{90}$ and $R_{50}$ have the advantage that they do not require a fitting, which can come with an error, so are more trustworthy in using them as scaling factors of the radii. $R_{90}$ has slightly more stable time evolution but all the results are qualitatively equivalent if we use $R_{50}$ instead. 

Additional quantities include the mean stellar age at $z=0$, and information about the stellar kinematics such as the mean stellar velocity and velocity dispersion.
We also measure the A2$(r)$ coefficient of the Fourier decomposition of the planar ($x$-$y$) stellar surface density (its maximum value at $z=0$ reported in Table \ref{proptable}). This is a measure of the strength of the non-axisymmetry at any given radius, and its peak value is usually taken as the strength of the bar in the galaxy. In Fig. \ref{barStrength}, we show in more detail the radial profiles of the A2 coefficient for each halo at different selected lookback times. Some of our halos have a very strong bar for most of their lifetime with values of ${\rm A2} > 0.3$. 

On top of the global disk properties, we also track the properties of individual stellar particles at any given snapshot which include their masses, positions, velocities, ages and metal content and we also have information about the birth radius of each star which is a useful quantity in this study. 

\begin{figure*}
    \centering
    \includegraphics[width=\linewidth]{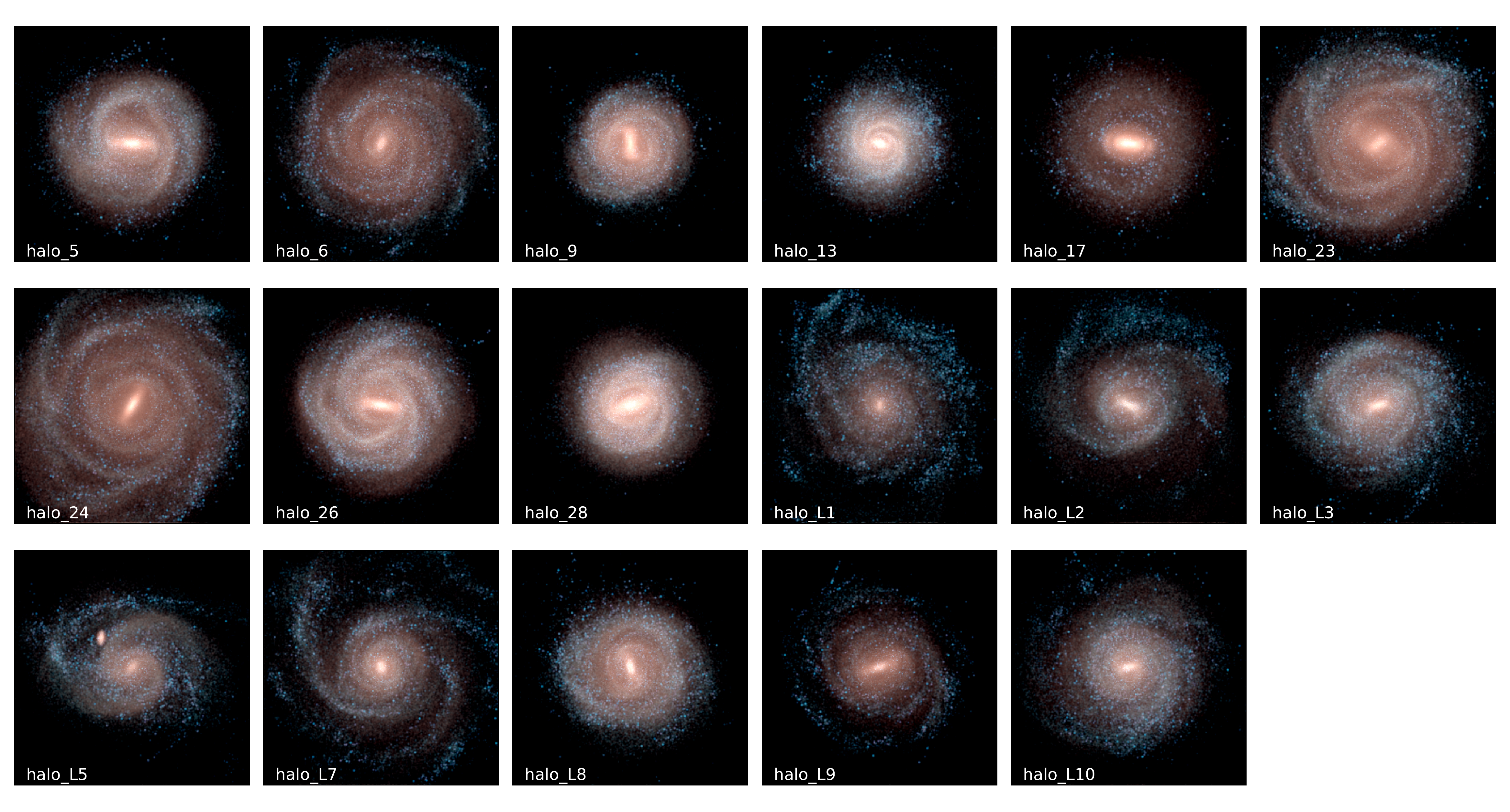}
    \caption{Stellar projections at $z=0$. The images are composites of the $r$, $g$ , and $i$ filters, and are rotated into the $x$-$y$ plane of the disk. The extent of all panels is 25$\times$25 kpc.}
    \label{stellarProj}
\end{figure*}

\begin{figure*}
    \centering
    \includegraphics[trim=2cm 2cm 2cm 0cm,width=\linewidth]{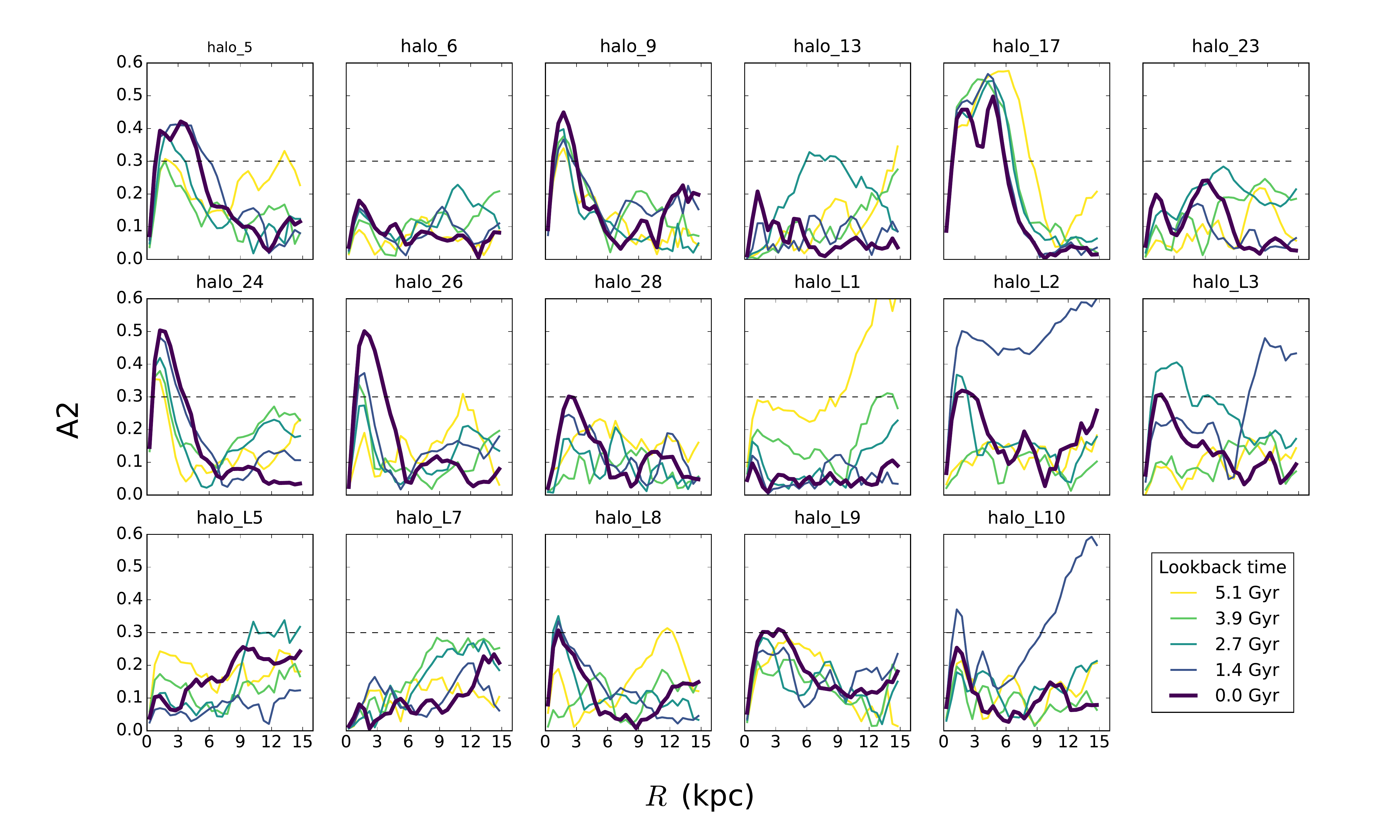}
    \caption{Measurements of the A2 coefficient in different radii as an indicator of bar strength for our systems at several different lookback times (shown with the different color curves). The bold purple line is the measured at $z=0$. A strong bar is considered to be present when A2 has values above 0.3 in the inner radii.}
    \label{barStrength}
\end{figure*}

In Fig. \ref{stellarProj} we present stellar projections at $z=0$ where the radial extent of each disk, the presence or not of the bar and the presence of spiral arms can be visually examined. The colours are composites of the \textit{g,r} and \textit{i} filters.

\section{Migration from the birth radius}

Similar to previous stellar migration studies, we look at the change in the galactocentric radius of stars between their birth time and the final snapshot of the simulation at $z=0$. We select all the stars in the disk at $z=0$, under the condition that they (a) have a circularity parameter of $\epsilon > 0.7$, to probe the cold stellar disk and exclude the bulge component, (b) are within 2 kpc of the galactic plane, and (c) are within 20 kpc from the centre of the galaxy. Furthermore, we only include stars that have been born in the main halo and not those that are accreted from other systems. 

\subsection{Overall changes in radius}

\begin{figure}
    \centering
    \includegraphics[trim=1cm 1cm 1cm 0cm,width=\linewidth]{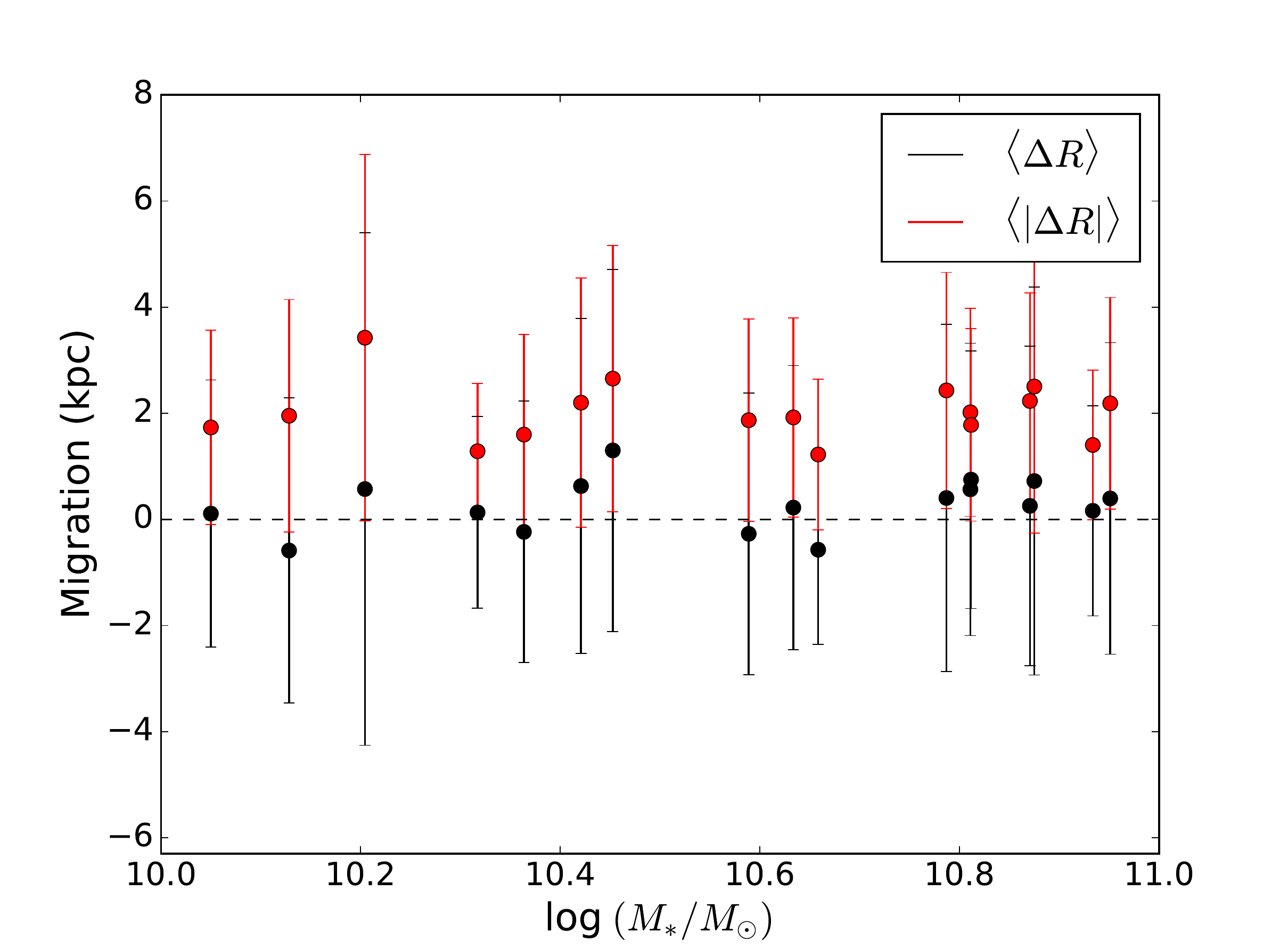}
    \caption{Mean change between the birth radius and the galactocentric radius at $z=0$ , $\Delta R = R_{z=0} - R_{\textnormal{birth}}$, for all disk stars and for all the systems in our sample plotted against the stellar mass at $z=0$. The black points show the average of the difference and the red the absolute value of the same quantity. The mean migration is very close to zero for all the systems.}
    \label{totalMigr}
\end{figure}

A first direct measurement from the simulation data is the computation of the overall change in radius, $\Delta R = R_{z=0} - R_{\textnormal{birth}}$, for each star. Fig. \ref{totalMigr} shows that, if we take the average of this value, $\langle{}\Delta{}R\rangle{}$, for all the stars selected as part of the disk, we find that in all our systems we get values that are very close to 0. In the same figure, we show the average of the \textit{absolute} value of $\Delta R$, $\langle{}|\Delta{}R|\rangle{}$, which for all our systems has values of around 1-3 kpc. In contrast to our findings, \cite{ElBadry16} find consistently positive values for $\langle{} \Delta R \rangle{}$ as well as higher values for $\langle{}|\Delta{}R|\rangle{}$ in similar plots. It should be noted however that they probe a very different mass range in their study and a different mechanism of migration due to stellar feedback.

\begin{figure}
    \centering
    \includegraphics[trim=1cm 1cm 1cm 0cm,width=\linewidth]{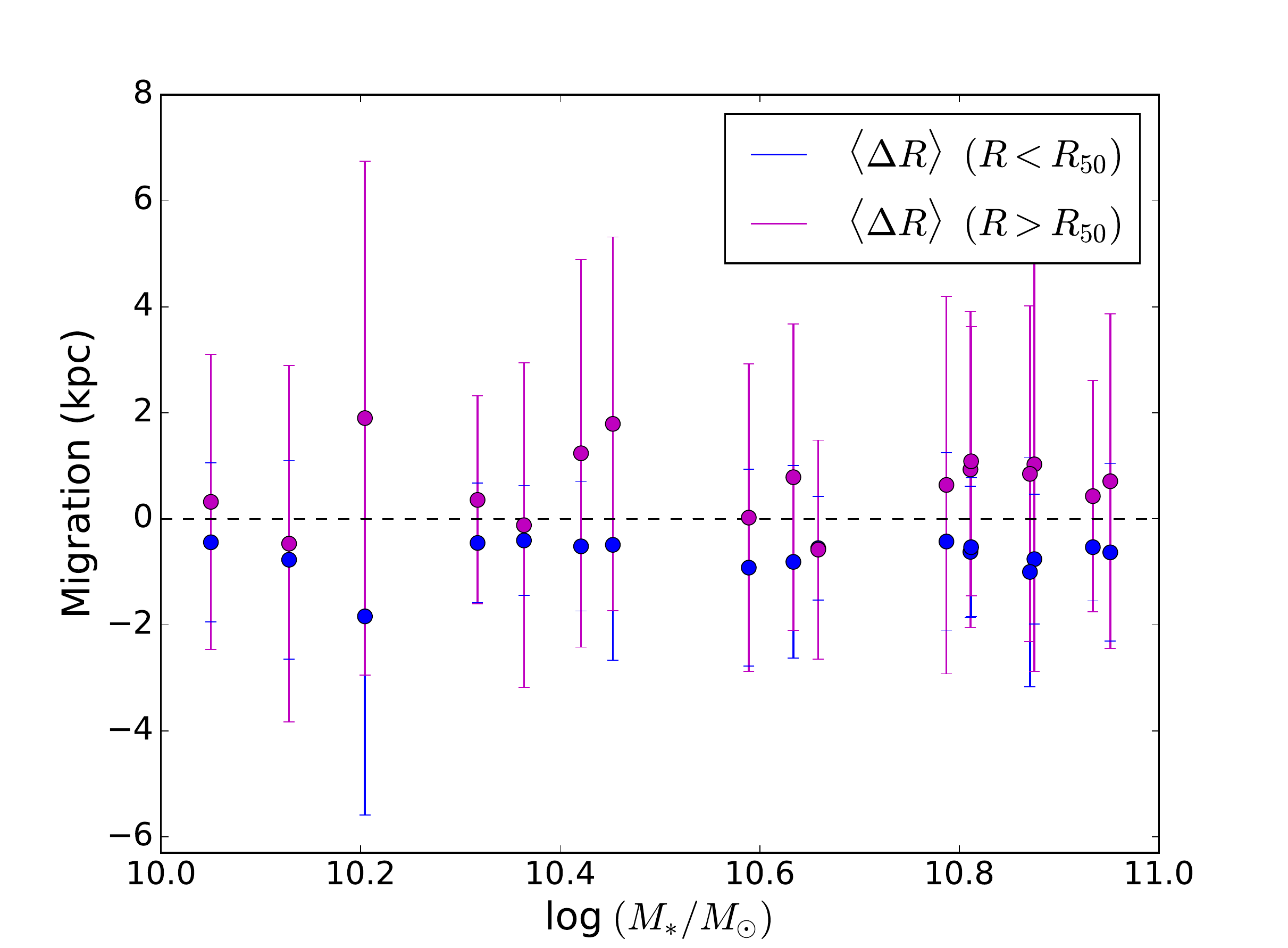}
    \caption{As Fig. \ref{totalMigr}, but for the mean migration for the stars within (blue) or outside (magenta) the half mass radius of the galaxy at redshift $z=0$. The former show on average negative values for most systems, indicating that their birth radius was in an outer region, whereas the latter have the opposite trend.}
    \label{totalMigrR50}
\end{figure}

In Fig. \ref{totalMigrR50} we compute the same average but instead selecting only stars that at $z=0$ are either inside or outside the half mass radius of the galaxy. We find that stars that are in the outer radii at $z=0$ have on average positive $\Delta R$, meaning that they have migrated outwards during their lifetime. The opposite is true for the stars that are within $R_{50}$ by $z=0$, which have a mean inwards migration.

Figs. \ref{totalMigr} and \ref{totalMigrR50} illustrate the combined effect of stars migrating both inwards and outwards within the disk, with a mean absolute migration scale in the range of a few kpc. This indicates significant mixing of material with different properties from the exchange of stars from inner and outer regions. We also see that there are no variations with the stellar mass of the system in our narrow mass range. 

\begin{figure*}
    \centering
    \includegraphics[trim=1cm 2cm 1cm 0cm,width=\textwidth]{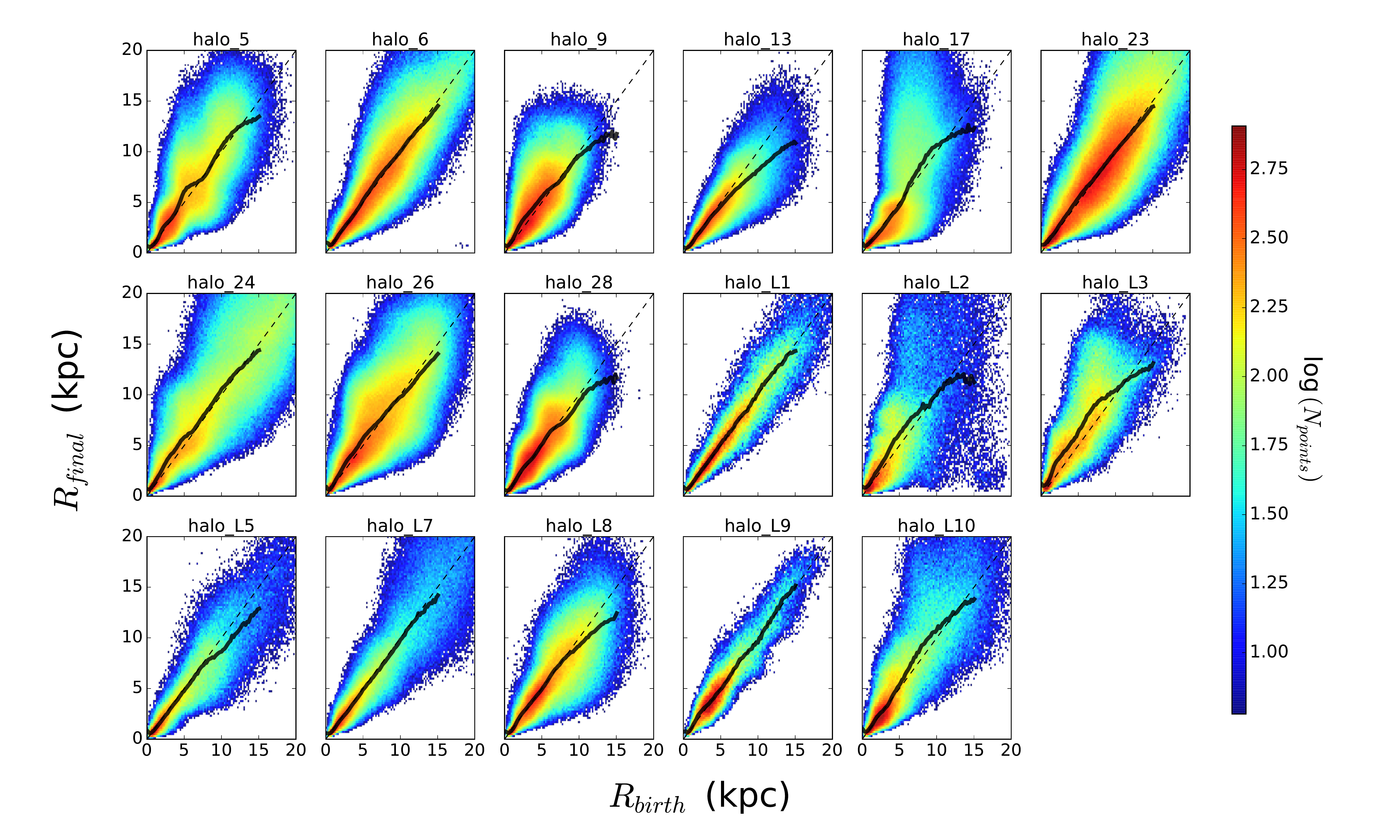}
    \caption{Birth radius versus final radius at $z=0$ for all the stellar particles with circular orbits in the simulated disks. The solid line represent the median $R_{\rm final}$ in bins of $R_{\rm birth}$ and the dashed line is the one-to-one line where $R_{\rm final} = R_{\rm birth}$. Most distributions appear symmetric around the one-to-one line but there are also cases like `halo\_17' (5th top row) or `halo\_9' (3rd top row), where there are several stars above the one-to-one line, born between 5-10 kpc but having migrated outwards to 10-15 kpc.}
    \label{rbirth_rfin}
\end{figure*}

We further look into the differences between the birth and final radii of stars by comparing directly the birth and final radii for stellar particles in each of our disks separately. In Fig. \ref{rbirth_rfin} we plot these two quantities against each other for the 17 indivdual systems. We find that most of the distributions are reasonably symmetric around the one-to-one lines, confirming the presence of both inwards and outwards moving migrators in roughly equal numbers, as also confirmed by the median line. However, some systems, such as `halo\_5', `halo\_9', and `halo\_17', show an excess of positive migrators (above the one to one line) for stars that have been born between 5-15 kpc. These can be identified as systems that have a strong bar in their centres which drives more significant migration. The other systems in our sample which exhibit a strong bar at $z=0$, such as `halo\_24' and `halo\_26' (see Fig.~\ref{barStrength}), also show evidence of a `bump' in their distribution in Fig.~\ref{rbirth_rfin} out to higher $R\sub{final}$.

We can decompose the information in Fig. \ref{rbirth_rfin} further by looking at the differences between birth and final radii in different radial bins and also for stars of different ages. To do this, we select stars in broad bins of birth radius and measure the distributions of their redshift zero radii. In addition, within each radial bin we split the stars based on their age at $z=0$, so that we can examine the strength of the migration for different stellar ages. We can compare our resulting distributions with the predictions from the study by \cite{Frankel18}, which models stellar migration around the solar radius as a diffusion process following a Gaussian function depending on one `migration strength' free parameter and the time $\tau$ after the birth of a star with a dependence that varies with the square root of $\tau$. 

\begin{figure*}
    \centering
    \includegraphics[trim=1cm 1cm 1cm 0cm,width=\textwidth]{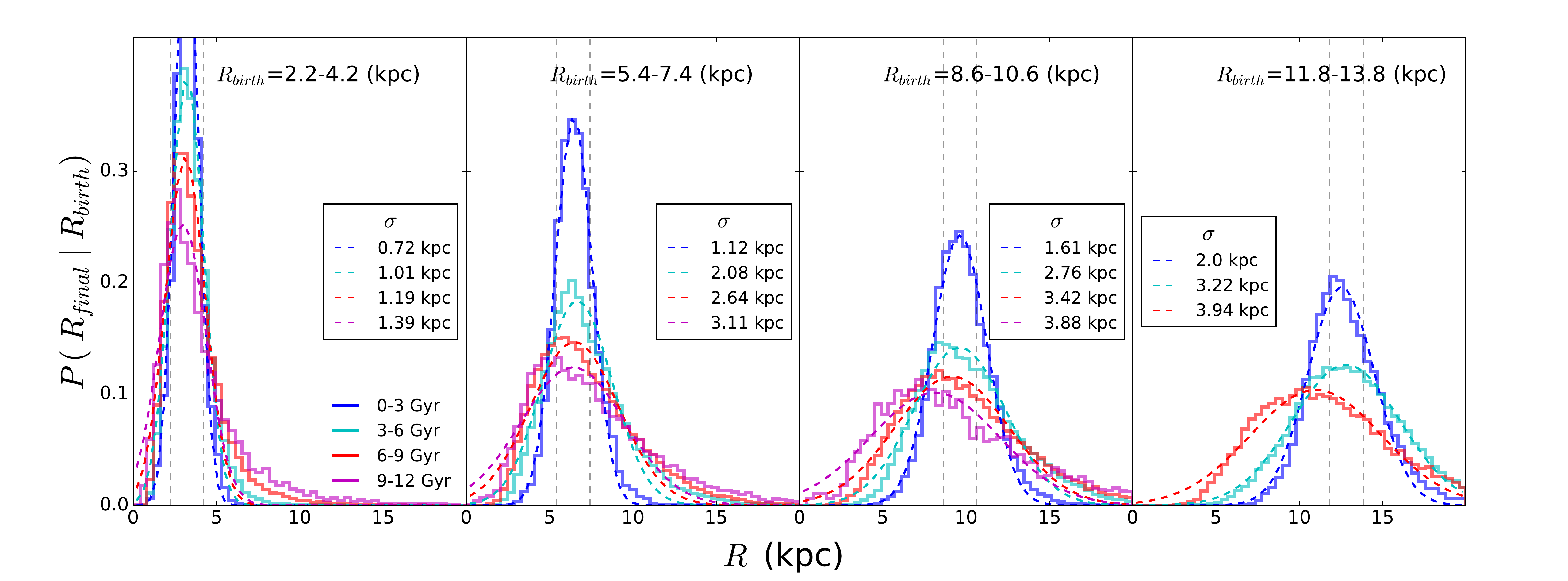}
    \caption{Probability distributions of final galactocentric radii for given birth radii bins, split by stellar age at $z=0$ for `halo\_6'. We overplot with the dashed lines the Gaussian fit to each histogram and in the legends we quote the 16-84 percentile of the histograms. We observe that the distributions become more extended with increasing stellar age as well as for stars that have been born in larger radii. The distributions are fairly symmetric and the peak is shifted from the centre of the selection region especially for the older populations. We stress that these Gaussian fits are meant to guide the eye, and are not used to quantify migration, which is instead measured by the 16-84 percentile range.}
    \label{halo6_hist}
\end{figure*}

In Fig. \ref{halo6_hist}, we show an application of this method for four distinct radial bins in one of our systems, `halo\_6', which is a typical Milky-Way-like disk with a quiet merger history and no strong bar in the centre. The resulting distributions are fairly symmetric and we find that their peaks can be fit reasonably well with a Gaussian function. However, the wings of the distributions, particularly in the inner part of the galaxy, are not fully described by a simple Gaussian distribution. Older stars appear to have diffused more from their birth radius, resulting in broader distributions, and additionally they show a larger shift away from the centre of their initial radial bin. In this particular case the shifts are inwards, as seen mainly in the two rightmost panels, but this is not the case for all systems and depends on the selection of the radial bin. Qualitatively this figure agrees with the similar Fig. 1 from \cite{Johnson21} both in terms of the widening of the histograms with age and the largest shift of the peak occurring for older populations.

In addition, this representative example agrees quite well with the \cite{Frankel18,Frankel20} model at radii around the middle of the stellar disk (\ie{}$\sim{}5.4-13.8$ kpc), in terms of the predicted spread of the histograms.This is encouraging because the Frankel models were fit to APOGEE DR12 data of Milky Way stars between galactocentric radii of 5 and 14 kpc. Although, we note that the small median shift we find in Auriga is not included in their functional form of migration. In order to present an average picture of what we observe in our whole sample, we measure the widths of these histograms in all cases for four stellar ages and four radial bins. We have checked that measuring either the 16-84 percentile or getting the variance of a Gaussian fit gives consistent results and we select the former to quantify the width of the histograms. We define this quantity, $\sigma\sub{migr}$, to be a measure of the migration strength.

\begin{figure}
    \centering
    \includegraphics[trim=3cm 0cm 3cm 0cm,width=\linewidth]{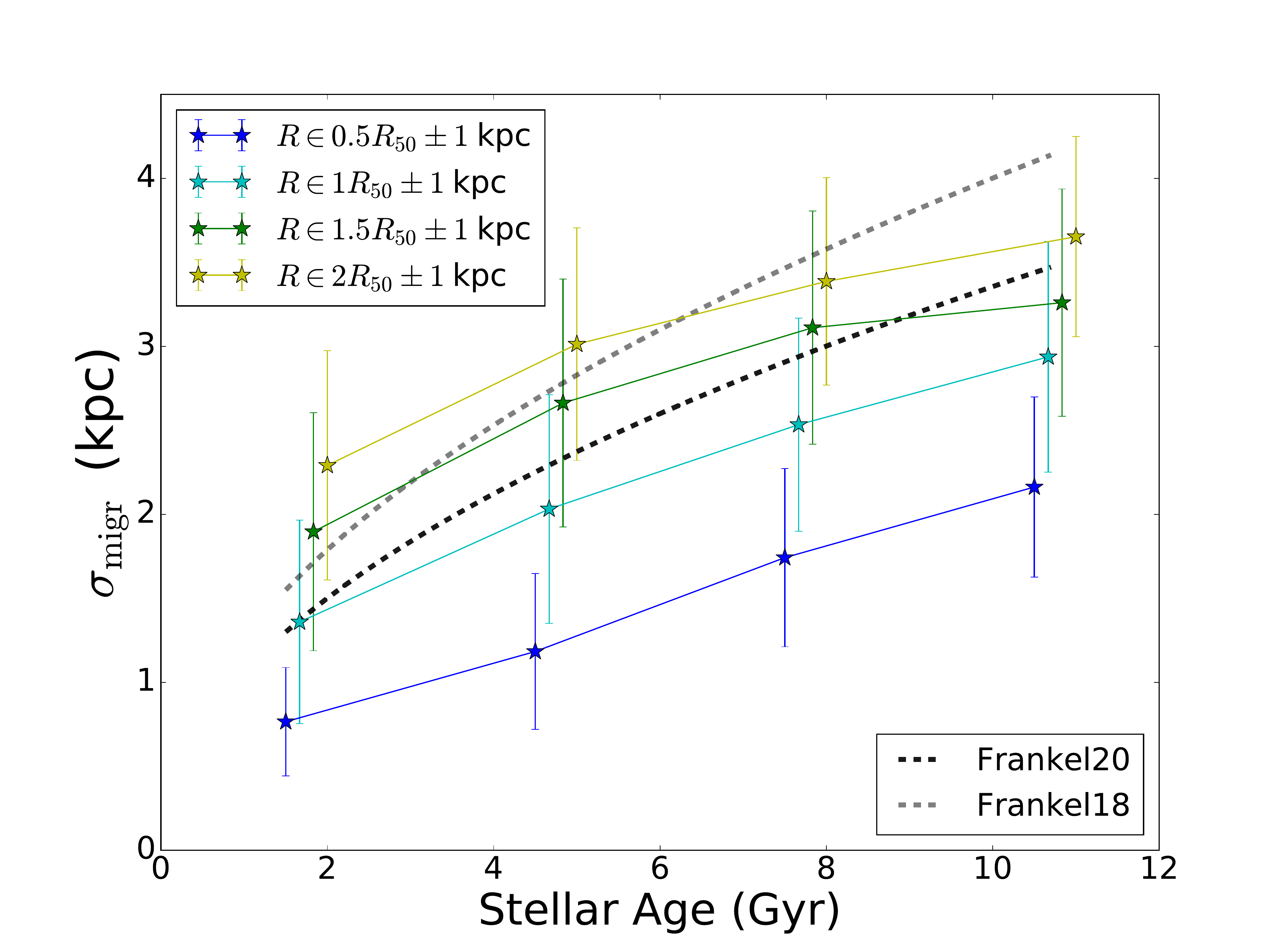}  
    \caption{Median values for the spread in stars from their initial positions (\ie{}migration strength) measured as the 16-84 percentile of the histograms in Fig.~\ref{halo6_hist}, as a function of their age at $z=0$. The solid curves show the median from all the halos in four radial bins of width 2 kpc centered on multiples of the stellar half mass radius, $R_{50}$. The dashed grey and black curves show two radially-independent models by \protect\cite{Frankel18,Frankel20}, for comparison.}
    \label{migr_age_medians}
\end{figure}

In Fig. \ref{migr_age_medians}, we show the average values for $\sigma\sub{migr}$ as a function of the stellar age, for each of the radial bins. We show the exact datapoints that were used to construct these curves in the Appendix Fig. \ref{migr_age}. For this figure, we have selected the radial bins to be normalised by the half-mass radius of each disk, thus accounting for the variation in disk size across the sample. We consider the centres of the four bins as the radii at 0.5, 1 , 1.5 and 2 times $R_{50}$ and select all the stars which have been born within 1 kpc of these radii. We obtain the mean age dependence for these four radial bins, and compare these to the radially-independent models by \cite{Frankel18,Frankel20} which we overplot. The median age dependence appears shallower in Auriga (for each of the radial bins) than in the Frankel models. However, the range of $\sigma\sub{migr}$ values found for our radial bins with $R \gtrsim{} 1\,R_{50}$ is roughly consistent with the model of \cite{Frankel20}, which is tuned to Milky Way stars at similar radii. The slope predicted by the Frankel et al. models is noticeably steeper than our curves at the three outer radial rings.  In particular a power law fit to the $\sigma\sub{migr}$-age curves gives us values of 0.4 (cyan), 0.28 (green), 0.25 (yellow), compared to the square root age dependence of 0.5 employed in the models.
The median values that we find for $\sigma_{\rm migr}$ are in the range of 1-4$\,{\rm kpc}$ depending on the radial bin and the age of the stars, which connects well to the values of $\langle{} \Delta R \rangle{}$ from Fig. \ref{totalMigr} for the whole population of stars. This is also broadly consistent with the values reported by \citet{Verma21} derived from the stellar metallicity dispersion of forward-modelled mock data for cross-matched {\it Gaia}, APOGEE, and {\it Kepler} observations.

\subsection{Effect of migration on radial profiles of age and metallicity}
\begin{figure*}
    \centering
    \includegraphics[trim=2cm 1cm 2cm 0cm,width=\textwidth]{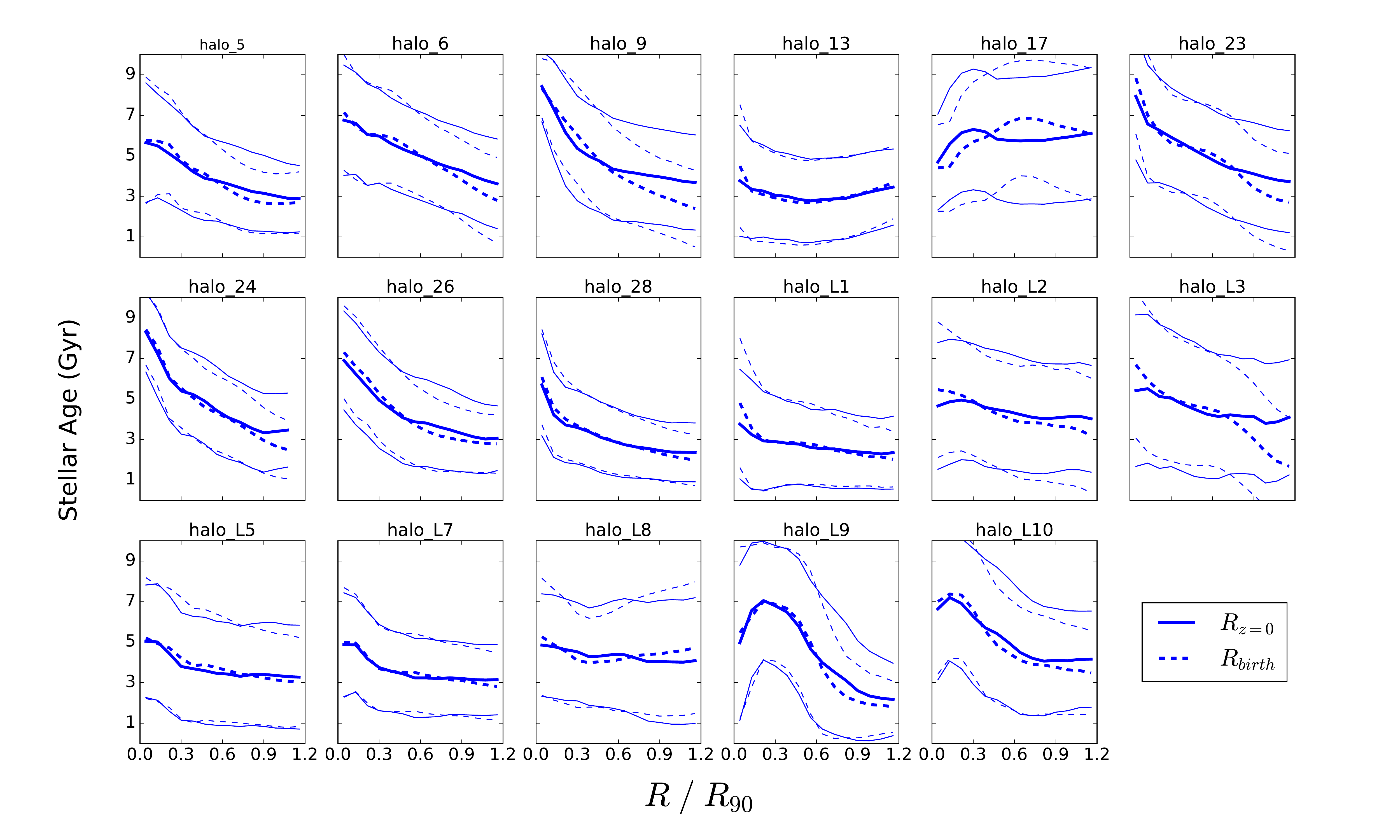}
    \caption{Mean stellar age profiles for the 17 halos in the simulation suite. The solid bold lines represent the true $z=0$ profiles whereas the dashed lines are the profiles that would have been obtained if the stars were located at their birth radii instead. With the thinner lines we show the variance of stellar ages around the mean.}
    \label{age_prof}
\end{figure*}

\begin{figure*}
    \centering
    \includegraphics[trim=2cm 1cm 2cm 0cm,width=\textwidth]{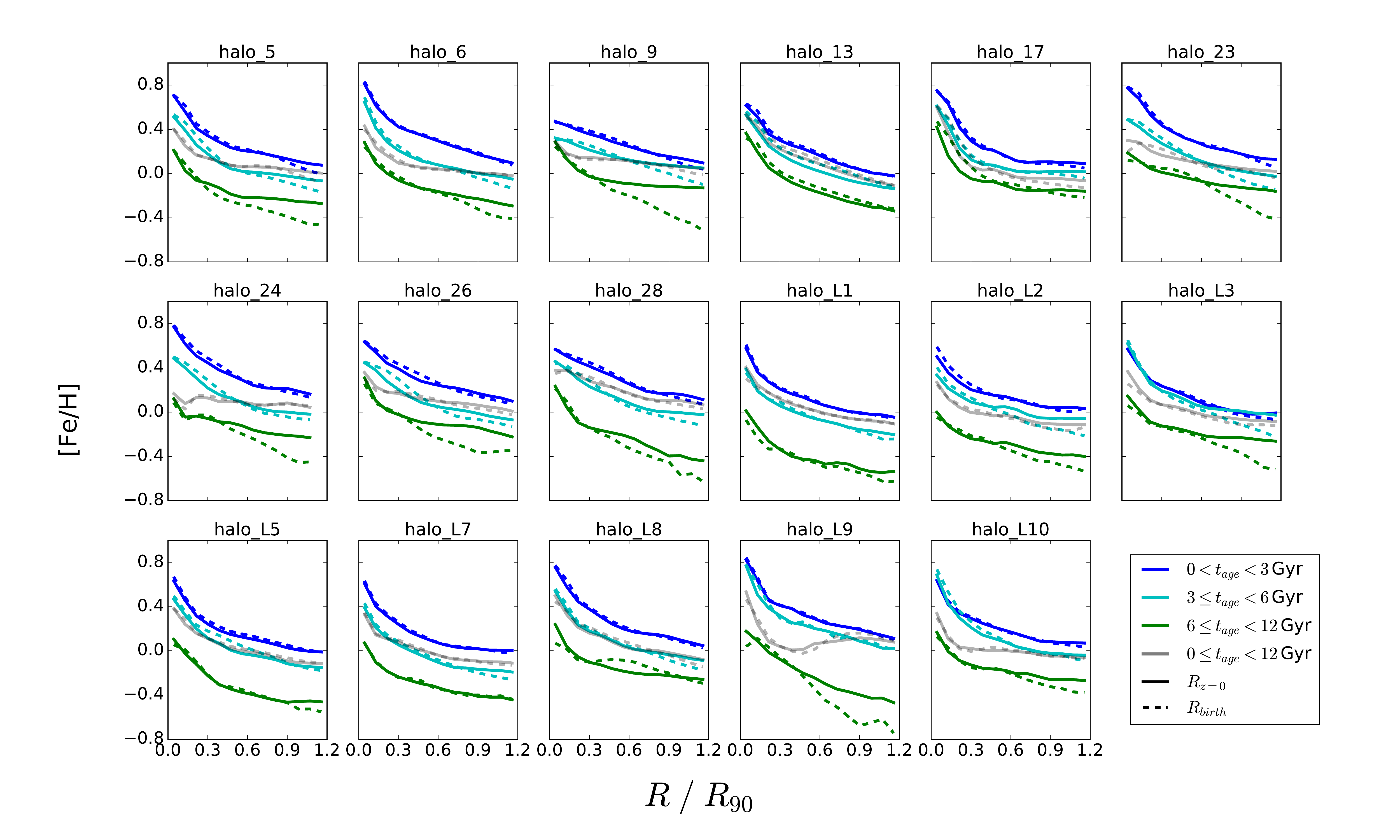}
    \caption{Radial metallicity profiles for the 17 halos in the simulation suite in three different stellar age bins. The solid lines represent the true $z=0$ profiles whereas the dashed lines give the profiles that would have been obtained if the stars were located at their birth radii instead. In many cases, for the older populations, there is considerable flattening of the true metallicity profile compared to what would be produced if there was no migration. This is similar to what is presented in \protect\cite{Minchev13}.}
    \label{metallicity_prof}
\end{figure*}

As a further step, we would like to investigate the effect of stellar migration on the metallicity and age profiles in Auriga. To do this, we plot the mean age and metallicity radial profiles for the stars in each of our simulated galaxies at $z=0$ in Figs. \ref{age_prof} and \ref{metallicity_prof} (solid lines). We use the $z=0$ galactocentric radii of the stars, normalised to $R_{90}$, which gives us the true profile observed in the simulation. The profiles are computing by averaging over the metallicities and ages of all stellar particles at a given bin of $R_{90}$. Additionally, we calculate the same profiles but assuming the birth radius of each star as its final radius, in other words simulating a scenario without any stellar migration (dashed lines in Fig.~\ref{age_prof}, \ref{metallicity_prof}). We then compare in each case the solid and dashed curves to evaluate the effect of stellar migration.

In the case of the mean age profiles (Fig.~\ref{age_prof}), we find differing results, depending on the halo, when comparing profiles with or without migration. In a number of halos, such as `halo\_9', `halo\_23' or `halo\_L3', migration of stars leads to the flattening of the profiles at outer radii, meaning that older stars have migrated outwards, increasing the mean age in those regions by $z=0$. On the other hand, there are cases such as `halo\_L7' or `halo\_13' where there is minimal difference between the two profiles, hinting towards very little stellar migration in these systems. There is no evident change in the overall scatter around the median. We already find an inherent scatter in the ages within each radial bin, suggesting that stars have formed at varying times at all radii, and migration does very little in further amplifying this spread.

In the case of the mean metallicity profiles (Fig.~\ref{metallicity_prof}), we choose to plot the solar-normalised iron abundance (in Auriga solar iron abundance is taken from \cite{Asplund09}), [Fe/H], which is a common indicator of a stellar population's metallicity. We calculate the total radial [Fe/H] profile for (a) all stars, and (b) stars that belong to different age bins. These are shown in Fig.~\ref{metallicity_prof}. Strikingly, we observe that the effect of migration appears not to be evident at all in the total profile, and barely noticeable for the younger stars in all the cases. However, for the older stars, in most cases, there is a flattening of the metallicity profile, due to more chemically enriched stars migrating to larger radii over cosmic time. This suggests that the lack of evolution in overall metallicity profiles is due to the outer disc being dominated by younger stars, which have not migrated as strongly as older populations.

\begin{figure}
    \centering
    \includegraphics[trim=1cm 2cm 1cm 0cm,width=\linewidth]{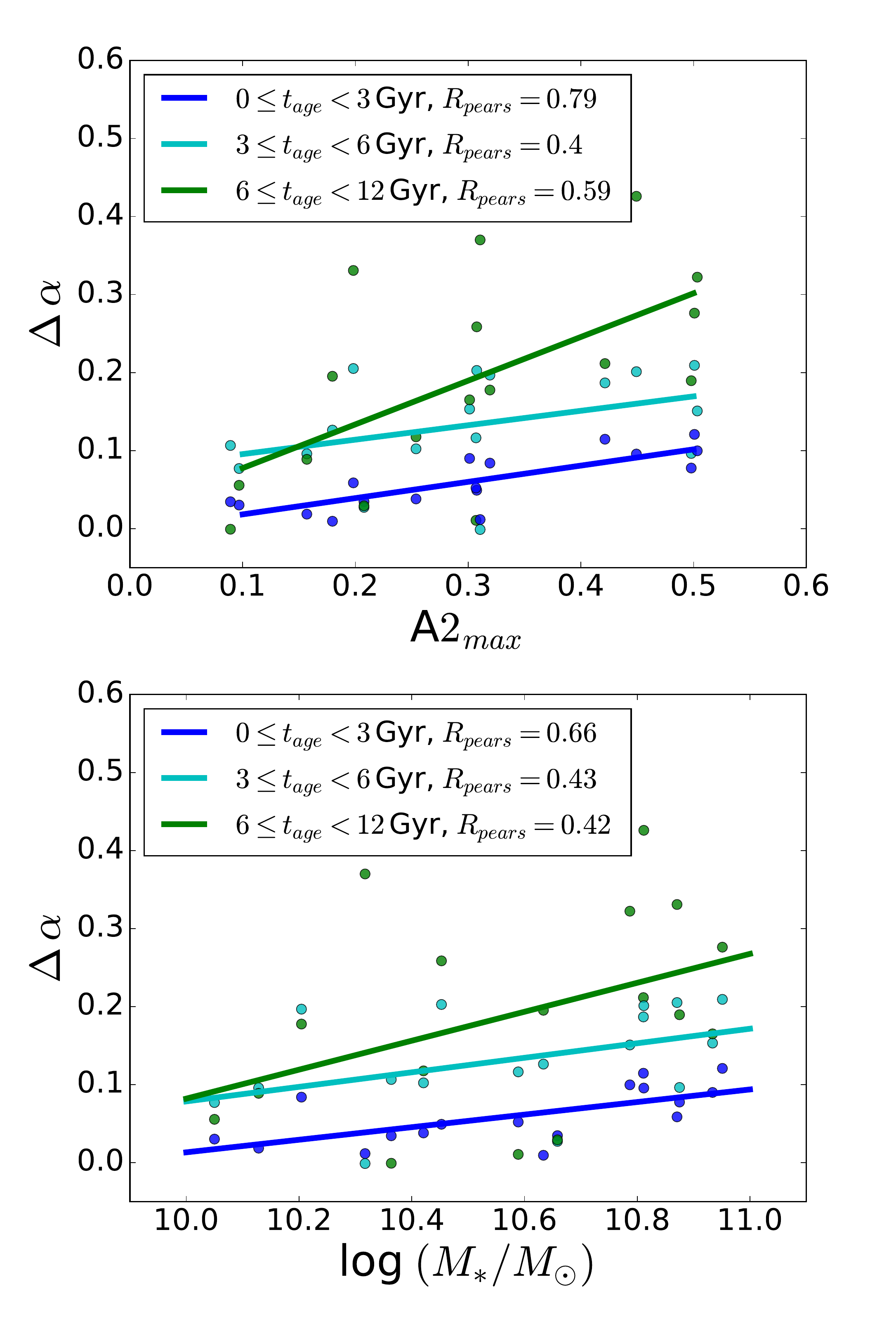}
    \caption{Change in the slope of the metallicity profiles ($\Delta \alpha$) presented in Fig.~\ref{metallicity_prof} measured for radii $R/R_{90}>0.2$. We show the same three stellar age bins and we plot the $\Delta \alpha$ against the maximum A2 coefficient (top) and stellar mass in the disk (bottom) at $z=0$. The lines are the linear fits to each data set, and we quote in the legend the Pearson correlation coefficient. There is a clear evolution with stellar age in the values of $\Delta \alpha$, and loose correlations of $\Delta \alpha$ with both A2 and the stellar mass are present.}
    \label{metallicity_slopes}
\end{figure}

To make a quantitative statement of this flattening, we fit both the true and the `birth-radii' profiles, excluding their core (\ie{}only for $R/R_{90}>0.2$), with a linear fit that gives a slope $\alpha$. We then measure the change in $\alpha$ between the true profile and the one without stellar migration, $\Delta \alpha = \alpha_{z=0} - \alpha_{\rm birth}$. Fig.~\ref{metallicity_slopes} shows the change in this outer slope for each different profile, plotted against the maximum A2 coefficient for the given disk (left panel) and its stellar mass (right panel). We examine the strength of the correlation between the change in the slope and these two quantities by calculating the Pearson correlation coefficient for each of the age sub-samples. A correlation is present with respect to both the stellar mass and the A2 coefficient despite the two quantities not being strongly correlated with each other.

The normalization of the best fit lines changes with the stellar age in the fashion we expect, with older stars showing higher $\Delta \alpha$ values. This reflects the clear flattening over time of metallicity profiles for older stellar populations. Nonetheless, there is also a shallower trend of increasing flattening with increasing bar strength and/or stellar mass for younger stellar populations. Indeed, the correlation appears stronger for the younger stars despite the small values of $\Delta \alpha$. But in all cases, the two-tailed p-values are sufficiently small to guarantee that the measured Pearson coefficients could not be drawn by a random uncorrelated sample. This demonstrates that migration is a more significant effect in (a) barred and (b) more massive Milky-Way-like galaxies in Auriga.

The results of this section highlight the advantage of studying a number of disk galaxies with different properties since the importance of stellar migration is not uniform for all the systems and diverging conclusions could be drawn if each system was to be studied individually. Furthermore, the decomposition of the metallicity profiles into age bins shows that migration does not leave the same imprint for stars of different age, and computing only the change in the \textit{overall} metallicity profile would hide the fact that stellar migration is occurring.

\section{Snapshot-to-snapshot migration}

The analysis carried-out above gives us a view of the total migration that has happened over the whole lifetime of the disk, but does not necessarily show us how migration evolves on shorter timescales. We are interested in this information as we are looking for an implementation of stellar migration for semi-analytic models of galaxy evolution where the model is updated on timesteps with a typical length of order $10$ Myr. Therefore, in this section we analyze the changes in the position of stars that happen between two consecutive simulation snapshots.

For each stellar particle we compute both the galactocentric radius and the guiding centre so that we can analyse the effect of each on the `churning' and `blurring' processes. A change in the galactocentric radius can result from both mechanisms but the changes in the guiding centres are solely due to torques exerted on the stars and directly relate to the `churning' process. The galactocentric radii are inferred directly from the simulation output as the distance between the position of a star and the centre of the disk. The guiding centres are computed using the information of the orbital angular momentum of a star, $l_{z,c}=RV_c$,  and interpolating this value to the rotation curve of the galaxy. From now on we refer to any quantity associated with guiding centres with a subscript `g'. Similar to the $R_{z=0} - R\sub{birth}$ calculations, we can look at initial, $R\sub{i}$ (or $R\sub{g,i}$), and final $R\sub{f}$ (or $R\sub{g,f}$), radii for the stellar particles, but in the subsequent analysis they are referring to any two snapshots in the simulation. 

\begin{figure*}
    \centering
    \includegraphics[trim=3cm 1cm 3cm 0cm,width=\textwidth]{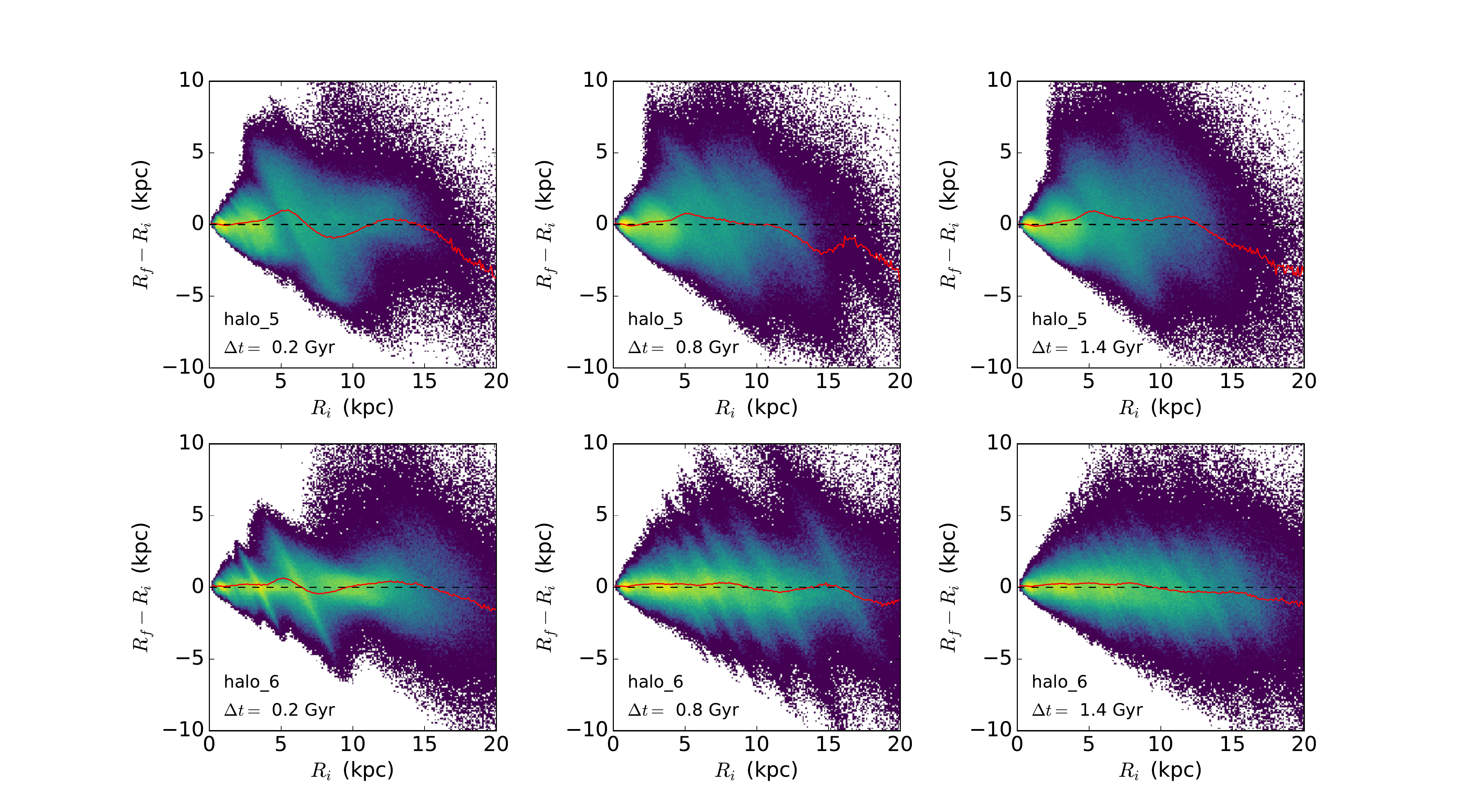}
    \caption{Change in galactocentric radius of the stars indentified in two snapshots plotted against their initial radius. The top row is an example of a galaxy in our suite which has a strong bar (`halo\_5') for most of its lifetime, whereas in the bottom row the galaxy has developed no bar (`halo\_6'). From left to right, we show three cases where the initial snapshot is the same but a different final snapshot separated by $\Delta t$ is chosen each time. Despite the galaxy specific differences in both disks we find that most stars have changes in their galactocentric radii between -5 to 5 kpc no matter how how large $\Delta t$ is.}
    \label{dr_r}
\end{figure*}

\begin{figure*}
    \centering
    \includegraphics[trim=3cm 1cm 3cm 0cm,width=\textwidth]{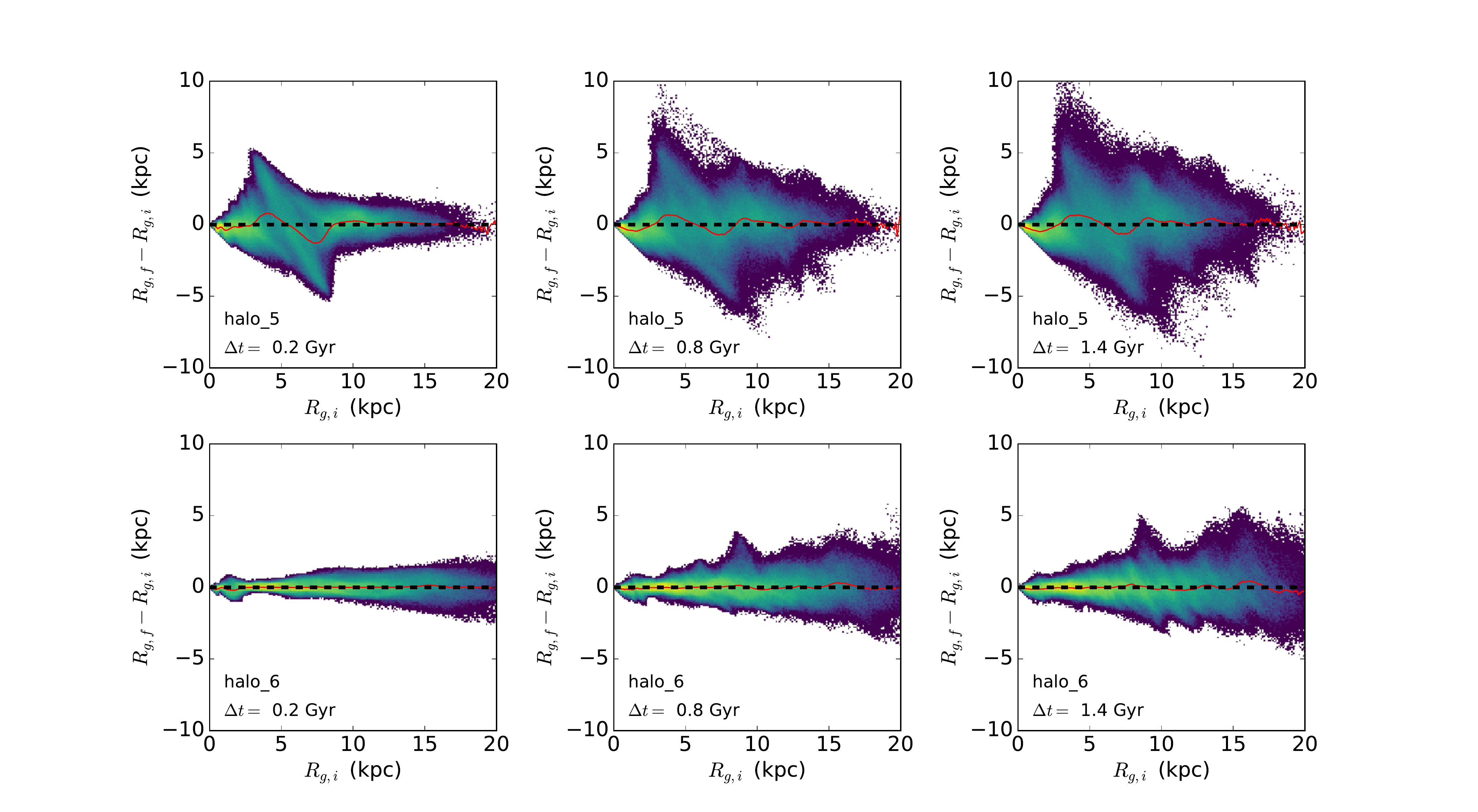}
    \caption{Similar as in Fig.~\ref{dr_r} but for the guiding centres of the stars. The same two galaxies are shown for the same snapshots. Here the difference between the barred and unbarred system is evident with the former having more extreme stellar migration, especially for stars located between 5-10 kpc. We also observe a widening of the migration pattern with time (from left to right).}
    \label{drg_rg}
\end{figure*}

Firstly, we directly estimate the difference between the initial and final radii (guiding or galactocentric) for our different halos and at a pair of snapshots. In Fig.~\ref{dr_r} we plot this difference in galactocentric radii $\Delta R = R\sub{f} - R\sub{i}$ against the initial galactocentric radius $R\sub{i}$, and similarly in Fig.~\ref{drg_rg} the difference in guiding centres $\Delta R = R\sub{g,f} - R\sub{g,i}$ against the initial guiding centre $R\sub{g,i}$. These figures are analogous to similar plots shown in previous studies such as \citep{Minchev12a, Minchev10} (showing change in angular momentum $\Delta L$ in the y-axis) or \cite{Halle18} (showing both $\Delta R$ and $\Delta R_g$). We present examples of two characteristic halos from our sample; one develops a strong bar during its lifetime (`halo\_5') and the other does not at any point in time (`halo\_6'). We show in both of these figures three different time intervals between the two snapshots, where the initial snapshot is always the same, at lookback time of 1 Gyr, and the final differs by the given $\Delta t$, in this case with values of 200 Myr, 800 Myr and 1.4 Gyr.

In the case of the galactocentric radii (Fig. \ref{dr_r}), the resulting patterns may not be identical between the two halos but they show similarities in the overall spread of amplitudes in the $y$-axis as well as little evolution with increasing $\Delta t$. The particular halo that we choose, `halo\_6', has a well-developed spiral arm pattern, which manifests in the $\Delta R - R_i$ plot as a series of diagonal regions of stronger migration around the location of the over-densities. It is interesting to observe that this effect becomes less pronounced if we look at snapshots that are separated by a longer timestep as it is most likely smoothed out by the longer time averaging and the transient nature of spiral arms \citep{Sellwood02,Grand12,Grand12b,Baba13}.  

Regarding the guiding centres (Fig. \ref{drg_rg}), there is a clear distinction between the two galaxies. In the case of the barred galaxy (`halo\_5'), there is an extended ridge of more strongly migrated (inwards and outwards) stars, the location of which matches with the co-rotation radius of the bar. Stars in this region have migrated up to 5 kpc outwards or inwards whereas in the other radii they are constrained within 2 kpc. The same pattern for a system without a bar (`halo\_6') has no distinct features, and we can only observe that the values for the migration are slightly larger in outer radii compared to the inner ones, although overall most stars appear to migrate less than 1~kpc. What is common in both halos is the widening of the patterns in the $y$-axis as we allow more time between the snapshots, indicating a process that is time (or timestep) dependent. 

These examples are representative of the behaviour that systems with or without a bar develop. Each halo in our sample could be studied on its own to get a much deeper understanding of each individual object, but we are more interested in this work in an average description of what we observe in our 17 systems. Thus, we would like to arrive at a formulation that describes, as generally as possible, the variations that we see in the $y$-axis of these plots for the galactocentric radii and the guiding centres.

\subsection{Stellar migration at different radii}

Our aim is to describe the effect of stellar migration at different radii and for this reason we perform a radial ring analysis. We split each galactic disk into a series of concentric annuli (rings) in the $x$-$y$ spatial plane extending out to 15~kpc in radius and 2~kpc above and below the disk plane in the $z$-direction. For the vast majority of snapshots, this radial extent is sufficient to enclose the entire stellar content of the disk. We choose a number of 20 rings, based on the radius $R$, that are of equal width and are linearly spaced in order to have a statistically reasonable number of stellar particles for each ring. We have tested for different height cuts and find that as long as we use a cut more than 1~kpc we obtain convergent and robust results.

Each stellar particle falls in a given ring based on its galactocentric radius. At the initial snapshot, \textit{n}, we record both its radius and compute its guiding centre. We then look for the same quantities at a subsequent snapshot, $n+m$. The selected stars have associated individual IDs which can be used to identify them at a later snapshot times in Auriga. Similarly to the analysis of birth-final radii above, we also impose a cut of $\epsilon > 0.7$ in the circularity parameter.

For each star, we compute the change in galactocentric radius as $\Delta R = R_{n+m} - R_n$, and the change in the guiding centre as $\Delta R\sub{g} = R_{\tn{g,}n+m} - R_{\tn{g,}n}$. We then create distributions of $\Delta R$ or $\Delta R_g$, associated with each ring, for a given halo and at a given initial snapshot \textit{n}. We show in Fig.~\ref{nmhist} examples of these histograms at two different radii for a particular halo. For a number of cases, the histograms can be approximately fit with a Gaussian function with the peak shifted from zero either to the negative (median inwards migration) or to the positive (median outwards migration), and are mostly symmetric around the median. However, since there are instances of heavily skewed distributions or distributions with broadened wings (as seen, for example, for `halo\_6' in Fig.~\ref{nmhist}), we refrain from universally fitting a Gaussian to extract the properties of each histogram. Instead, similar to the analysis of radial gas flows in Auriga presented in \cite{Okalidis21}, we quantify the resulting distribution by (a) its median, $\overline{\Delta r}$ and $\overline{\Delta r\sub{g}}$, and (b) its width (\ie{}radial spread), $w$ and $w\sub{g}$. We choose to measure the width as the value of the 16-84th percentile of the distribution that corresponds to 1$\sigma$ of a normal distribution. We carry out this process for different values of the initial snapshot $n$, starting from $n=150$, which corresponds to a lookback time of $t\sub{look}\sim 6$ Gyr in our simulations, up to the last available snapshot based on the value of \textit{m}. We do not look further back in time because we are interested in the regimes when our halos have formed a well-developed rotationally supported stellar disk, which may not be the case in some of our systems at earlier redshifts.

We repeat this analysis varying the value of \textit{m}, namely we use $m = 1$, 3, 5, 7, 10, and 15, to constrain the evolution of $w$, $w_g$ and $\overline{\Delta r}$, $\overline{\Delta r_g}$ with increasing snapshot spacing. The option of $m=1$ equates to about 60 Myr in our simulations and $m=15$ to a timespan of 1 Gyr. For the presentation of some results in the next sections we choose as a default the value of $m=10$, which corresponds to $\Delta t = 600-700$ Myr between two snapshots which is approximately 3 dynamical times at the Solar radius. We believe this is a sufficient period of time to approximate migration as a diffusive process. To summarize, we obtain data points for the four $w$, $w\sub{g}$, $\overline{\Delta r}$, and $\overline{\Delta r\sub{g}}$ quantities which are associated with a set of variables, the centre of the ring, the initial and final snapshot times and their difference and the specific halo that the ring belongs to ($R\sub{ring,j}$, $t_n$ ; $\Delta t=t_{m}-t_{n}$, halo$_k$).

Furthermore, for a given choice of \textit{m}, we analyse how migration depends on stellar age by binning stars into groups of 0-1 Gyr, 1-3 Gyr, 3-5 Gyr, and 5-8 Gyr based on their ages at the initial snapshot of selection, and applying the same analysis as above.


\subsection{Radial Profiles}
In this section we analyze how $w$, $w\sub{g}$, $\overline{\Delta r}$, and $\overline{\Delta r\sub{g}}$ vary according to radius. For each ring, we get a normalised value for the ring's mid-radius, $R\sub{sc} = R\sub{mid}/R_{90}$, where $R_{90}$ refers to the 90\% mass radius at the initial snapshot \textit{n}. In Fig.~\ref{rad_prof_snap} we show the normalised radial profiles for our four quantities that result from averaging the data from all our systems. In each panel the curves represent a different value of $\Delta t$ between the initial and final snapshot.

The first important result to take away from Fig.~\ref{rad_prof_snap} is that there are qualitative and quantitative differences in the radial profiles when using the galactocentric radii or the guiding centres as indicators of migration. In general, the values for $w$ and $\overline{\Delta r}$ are larger than the corresponding values for $w\sub{g}$ and $\overline{\Delta r\sub{g}}$ for the same time difference. In both cases, the width of the distributions increase as we move towards the outer rings, but in the case of the median shifts we find higher negative values (inflow) for $\overline{\Delta r}$ as we move outwards in the disk than we do for $\overline{\Delta r\sub{g}}$. It is worth noticing that $\overline{\Delta r\sub{g}}$ and $\overline{\Delta r}$ have almost exclusively negative values at all radii which corresponds to stars moving inwards on average in the given ring. That is simply the median of the histograms that comes out negative in most cases, but there is always a significant number of stars that have migrated outwards. This information is better captured in the values of $w$ and $w_g$. 

\subsection{Time interval dependence}

Concerning the dependence of the median shift and the spread on the snapshot spacing, we see a clear evolution in the radial profiles over all radii for $w_g$ and $\overline{\Delta r_g}$. As expected, if we allow more time between the two selected snapshots the width of the histograms is larger, that is to say stars appear to have diffused more strongly out from their initial positions. The respective evolution for $w$ is much more obscure and in particular the curve for the case where we allow our minimum timestep ($n+1$) does not follow the same trend as the other choices of $\Delta t$. Moreover, in the case of $\overline{\Delta r}$ it can be argued that there is no clear timestep dependence at all as the several mean curves for the different timesteps overlap significantly. 

\begin{figure*}
    \centering
    \includegraphics[trim=1cm 1cm 1cm 0cm,width=\textwidth]{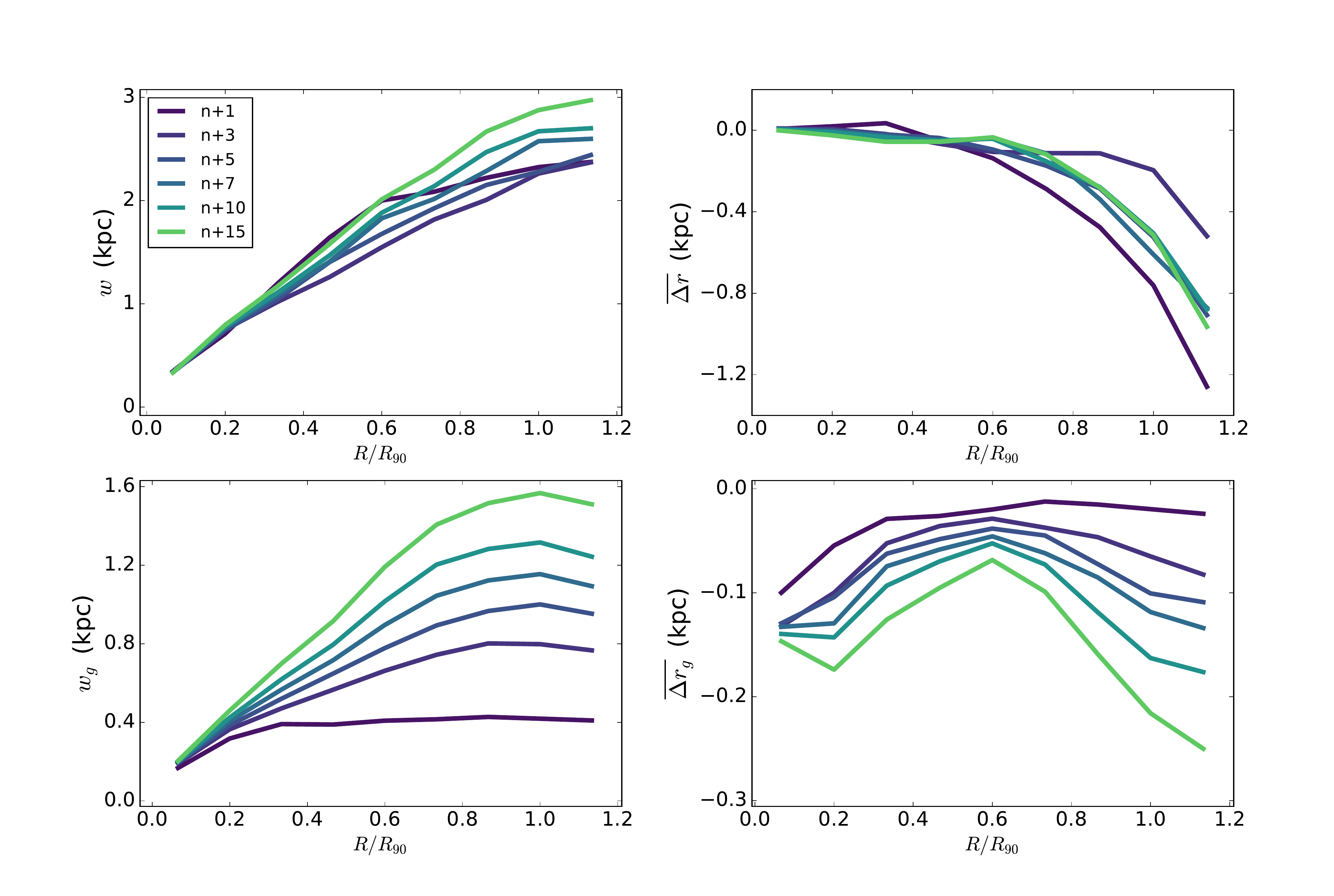}
    \caption{Mean radial profiles of the quantities that describe the distributions of $\Delta r$ and $\Delta r_g$ plotted in terms of the normalised ring radii. In the left panels we plot the spread of the histograms and in the right the median. The top panels are for galactocentric radii and the bottom ones for the guiding centres. Here we plot these profiles for different selections of snapshot spacings between the initial and final positions to show the evolution with $\Delta t$. The latter appears to be more pronounced and more clearly defined when changes in guiding centres are considered.}
    \label{rad_prof_snap}
\end{figure*}

If we assume that migration, as the change of the guiding centres, was a pure diffusion process, we would expect to recover a $w\sub{g} \sim \Delta t^{0.5}$ dependence. Therefore, following the approach used by \cite{Okalidis21}, we assume that there is a more general time dependence of the form $w \sim \Delta t^a$ and construct logarithmic plots of $w$ and $w\sub{g}$, with the aim to extract the slope, thus giving us a measurement of the actually realized exponent $a$. From Fig.~\ref{rad_prof_snap}, it is already evident that this time dependence is not the same in different radial scales. Thus, we split the data into three broad bins of normalised radius to determine if there is also a radial dependence of $a$.

\begin{figure}
    \centering
    \includegraphics[trim=1cm 1.5cm 1cm 0cm,width=\linewidth]{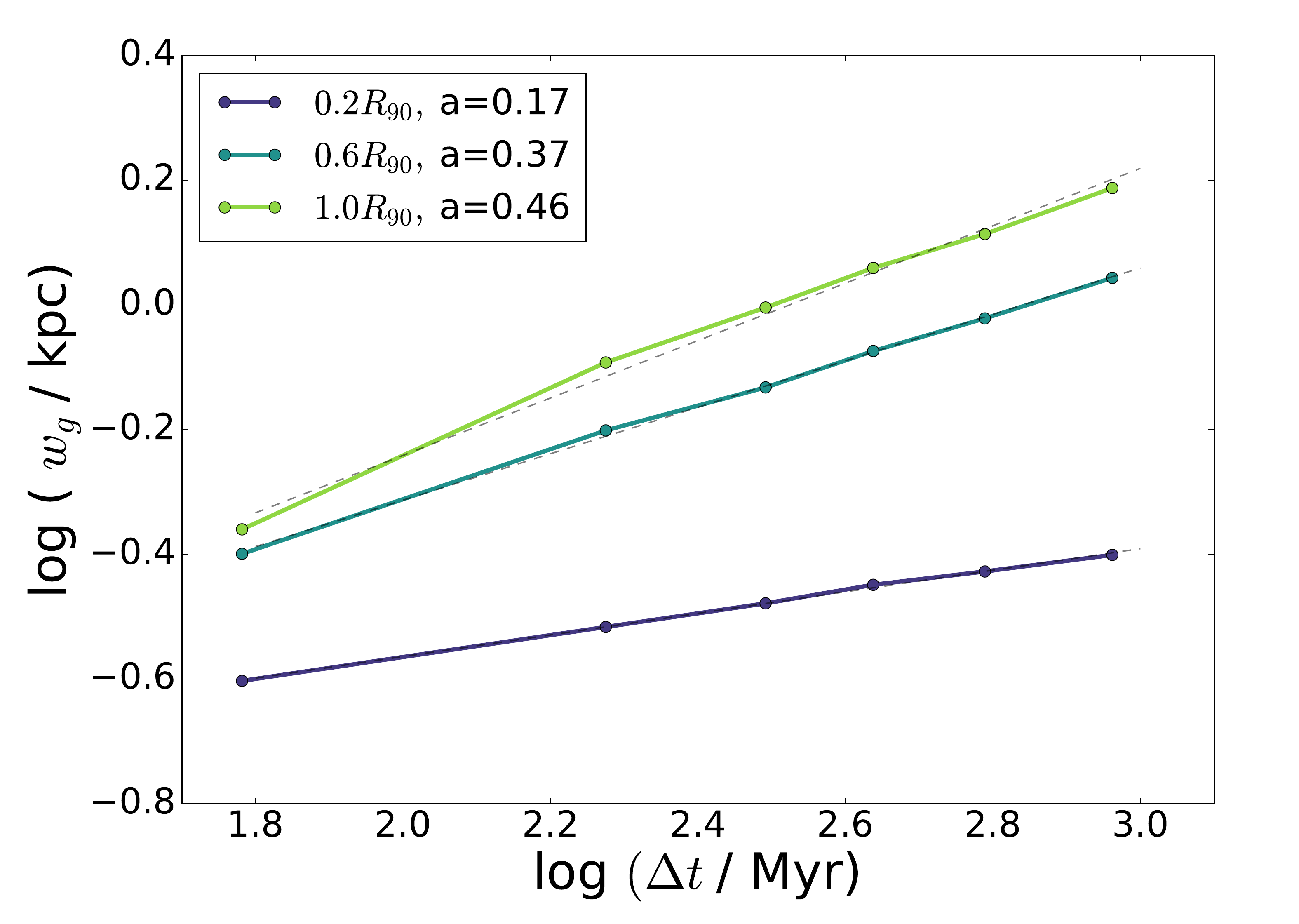}
    \caption{Logarithmic plot of the spread $w_{g}$ against the time interval $\Delta t$ based on the average of the combined data for all the halos. We show the trend in three different radial bins and observe a variation of the  time interval dependence with radius, with a stronger effect being found in the outer radii.}
    \label{wg_timefit_all}
\end{figure}

This is shown in Fig. \ref{wg_timefit_all} for the $w_{g}-\Delta t$ relation for the combined data for all the halos. In this case, we do not recover the $a=0.5$ exponent expected for pure diffusion in any radial bin, but there is a radius evolution that approaches this theoretical value at larger radii. We should not necessarily expect to get the theoretical slope for the average of all the halos, which may have significantly different evolutionary history and structural properties. So, we further calculate the exponents in each individual system. This is shown in the appendix Fig. \ref{wg_timefit}, from which it immediately becomes evident that there is a range of values for $a$ obtained for the different halos. Overall, in the innermost radial bin, the slope is significantly flatter in all cases. This appears reasonable because these central regions are often dominated by a bulge or a bar rather than featuring a very flat rotation dominated stellar disk and indeed there is a correlation in that strongly barred systems have a lower exponent for the inner ring. In contrast, at large radii, there are cases where the values of $a$ are reasonably close to the diffusion case (for example, in `halo\_6') and we do not see a strong correlation between the bar strength and the slope. 

\begin{figure}
    \centering
    \includegraphics[trim=1cm 1.5cm 1cm 0cm,width=\linewidth]{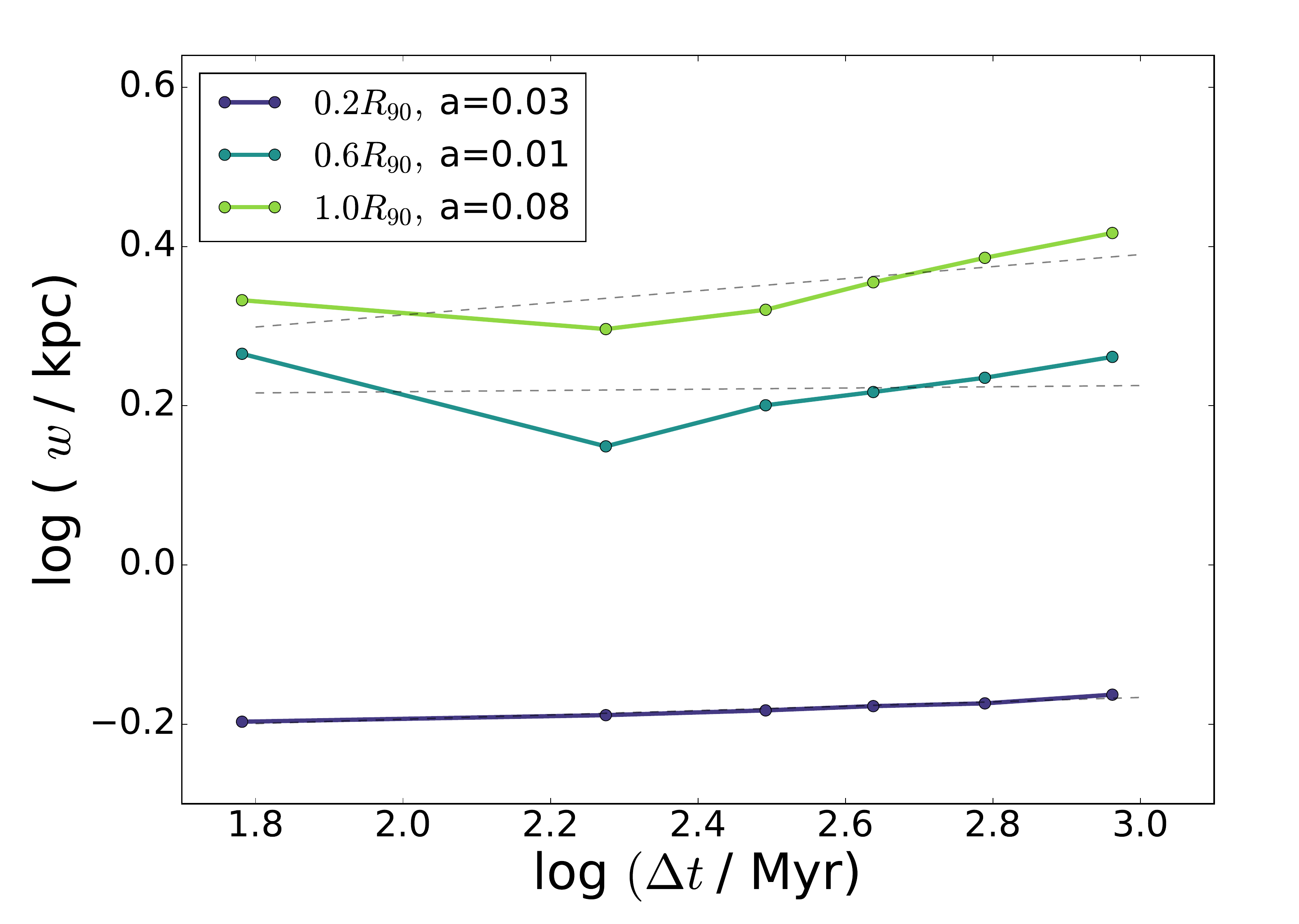}
    \caption{Logarithmic plot of the spread $w$ against the time interval $\Delta t$ based on the average of the combined data for all the halos. We show the trend in three different radial bins.}
    \label{w_timefit_all}
\end{figure}

The same figure, Fig. \ref{w_timefit_all}, for the $w -\Delta t$ confirms what was found before, that the shortest timestep does not follow the same linear trend in the log-log plot resulting in a bad fit which yields a very low value for the exponent $a$. On top of that, $a$ varies significantly if we calculate it individually for each halo and is in most cases much flatter than the diffusion value of 0.5. It must also be noted that although it looks like in the combined dataset that the rest of the points follow a linear trend, this does not appear in many of the individual systems. Therefore, from our data, we cannot conclude the presence of a robust time dependence such as we found in the case of guiding centre quantities.

There is also a clear timestep dependence in the median shift in guiding centre radius, $\overline{\Delta r\sub{g}}$. Although it can also be quantified at the different radial bins, we find that it varies much more significantly from halo to halo, and in many individual cases there is no discernible time interval dependence at all. 

All in all we cannot safely suggest that the migration process is purely diffusive in the average of our sampled disks, however it can be described as such in certain individual systems. Moreover, it is only when we compute the changes in the guiding centres that we can retrieve a timestep dependence that is similar to a diffusive process, since the changes in galactocentric radii seem to be following a much flatter time evolution both in the average and the individual systems.

\subsection{Bar and age dependence}

\begin{figure*}
    \centering
    \includegraphics[trim=1cm 1cm 1cm 0cm,width=\textwidth]{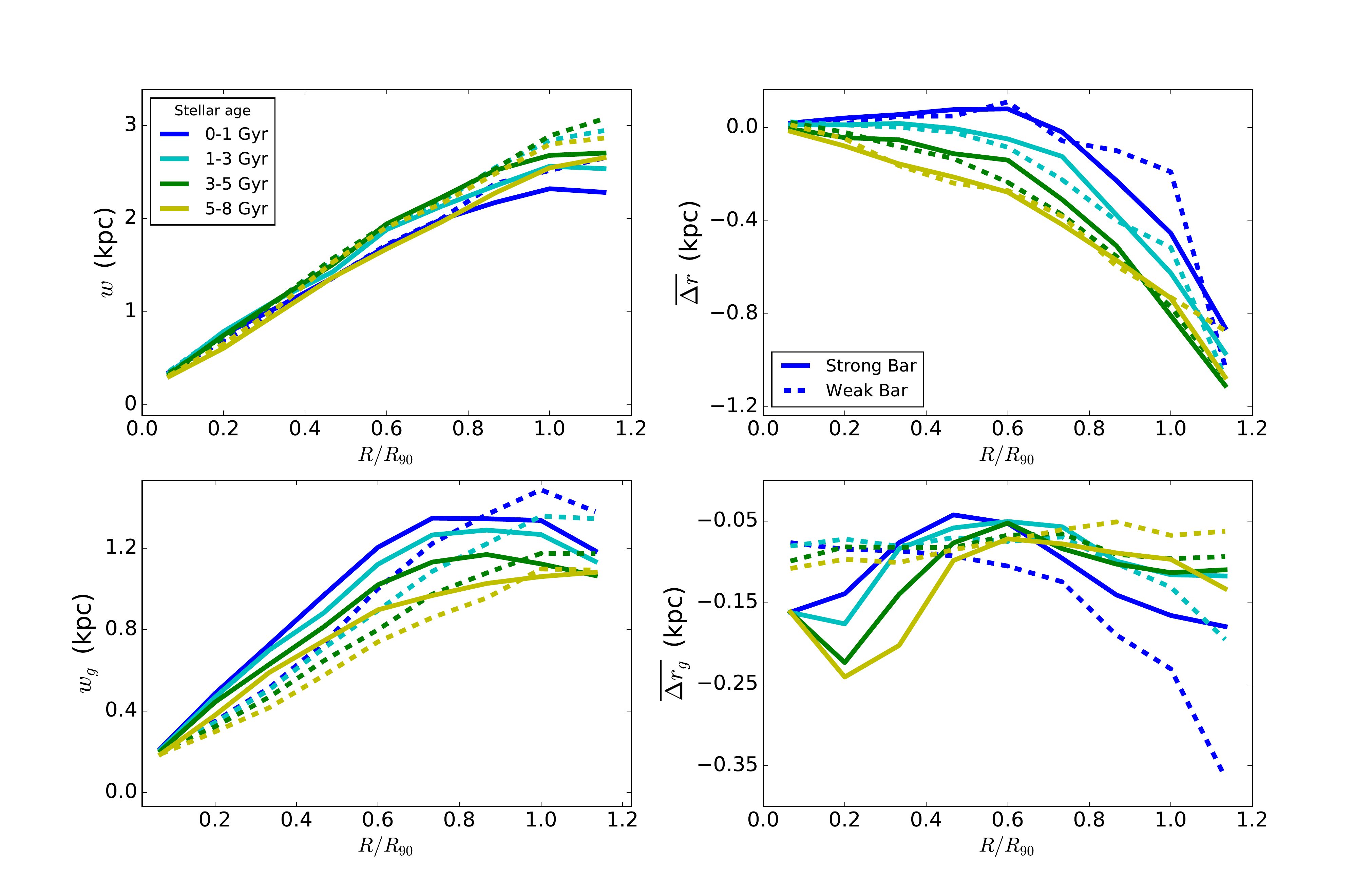}
    \caption{Similar profiles as in Fig.~\ref{rad_prof_snap} for a fixed snapshot spacing of $n+10$, corresponding to 600-700 Myr. We split the sample between the halos that develop a strong bar, $A_{2,z=0} > 0.3$, for most of their evolutionary history (solid lines) and those that have a weaker or no bar, $A_{2,z=0} < 0.3$, (dashed). The different colored curves in each panel show the quantities as obtained by using only stars belonging to different age bins (legend in top left panel) selected based on their age at the initial snapshot $n$.}
    \label{rad_prof_agebar}
\end{figure*}

In Fig.~\ref{rad_prof_agebar}, we plot the radial profiles for one fixed value of $\Delta t$ ($n+10$), in order to study the effect of the presence of a bar in the disk, as well as any differences between stars of different ages. We split our sample in two sub-samples of strongly-barred and weakly-barred systems, based on the maximum A2 value at redshift zero, with each sub-sample then having 10 and 7 galaxies, respectively, if we take 0.3 as the separating value. 

We find that, in the case of the galactocentric-radii-based quantities $\overline{\Delta r}$ and $w$, on average, the radial profiles are consistent with each other between the strongly-barred and weakly-barred galaxies, suggesting that the information conveyed by these two quantities cannot be directly associated with the presence or absence of a bar. This is reasonable as (1) $\Delta r$ is largely influenced by the exact position the star particle is captured at the particular snapshot time during its orbit, and (2) there are additional factors that can alter the galactocentric radius of a stellar particle. 

There is a greater distinction in the radial profiles for guiding-centre-based quantities. The barred galaxies have on average higher values of $w\sub{g}$ at all radii, with the difference being maximised at the central part of the disk around the co-rotation radius of the bar. This is more evident if we study the radial profiles for individual galaxies. It is not very pronounced when averaging all the barred systems because of the different strengths and radii of the bars so that we can have a partial cancelling of the bar effect by averaging a region which is dominated by the bar co-rotation in one halo and a region that is further in or out of the co-rotation radius in another. These profiles reflect what we observe, plotted in a different manner, in the $\Delta R\sub{g}-R\sub{g,i}$ (Fig. \ref{drg_rg}) and $\Delta R-R\sub{i}$ (Fig. \ref{dr_r}) plots. 

Concerning $\overline{\Delta r\sub{g}}$, there is significantly increased inward migration in the innermost regions of strongly-barred systems compared to weakly-barred systems, as shown in Fig.~\ref{rad_prof_agebar}. This feature appears in both the average profiles and each individual barred halo. The weakly-barred systems have very flat profiles for radii $\sim{}0.6\,R_{90}$ with less inward migration.

In Fig.~\ref{rad_prof_agebar}, we also explore the stellar age dependence of each of the quantities $w$, $w_g$, $\Delta r$, and $\Delta r_g$ by showing four different age bins. When measuring the total migration in the previous sections, there is a hint of an age dependence, with older stars experiencing larger migration (Fig. \ref{halo6_hist}, \ref{migr_age_medians}). This is not immediately reflected in the values of our measured quantities in the different age bins. The spread $w$ shows no evidence of a consistent age dependence with the different curves overlapping with each other. The corresponding spread $w\sub{g}$ has a clear age dependence with up to a 30 per cent difference between the youngest and oldest bin. This dependence is the opposite to what was found before for the total migration --  stars in the earlier stages of their lives appear to be more diffusive in terms of the changes in their guiding centres. We must stress though, that the age definitions in this and the previous sections are not identical since in section 3 we refer to the age of the star at $z=0$ whereas in this section it is the age of the star at a given snapshot. The effect itself can be explained on the basis that younger stars, having lower velocity dispersion, are more prone to be impacted by angular momentum exchanges that can happen in the disk \citep[e.g.][]{VeraCiro14}. The overall larger effect for the older stars when looking only at the difference between $z=0$ and the star's birth radius is merely due to the fact that the small instantaneous changes are adding up for a longer time. The median shifts, $\overline{\Delta r\sub{g}}$ and $\overline{\Delta r}$, show a similar age dependence, with older stars having more negative values, meaning that they are on average shifted more strongly inwards although the trend reverses in the outer rings for $\overline{\Delta r\sub{g}}$.

\subsection{Parametrization of stellar migration in Auriga}

Our goal in this section is to present a simple parametrization of the strength of radial migration in Auriga, which we have so far described either by $w$ and $\overline{\Delta r}$ (for changes in galactocentric radii) or $w_g$ and $\overline{\Delta r_g}$ (for changes in guiding centre radius). In the case of the galactocentric radii, as we saw in the previous sections, there are weak but quantifiable trends with the size of the timestep interval between the analyzed snapshots. However, there are no trends with respect to the presence or absence of a bar, and a dependence on the different stellar ages only for the quantity $\overline{\Delta r}$. For the guiding centre radii, the difference in the median shift radial profiles for the barred and unbarred populations, seen in Fig.~\ref{rad_prof_agebar}, suggests that separate parametrizations for them may be needed. Indeed, we do not recover the exact same timestep dependence for weakly and strongly barred galaxies, but for sake of simplicity in the parametrization and to increase the number statistics we opt to focus on the average of all the 17 halos. The dependence of some of the fit parameters in the presence of a bar is stated later in the text. In the previous section we presented variations of the time dependence that exist at different radii but for the purposes of a global parametrization we want to collectively describe the curves that we extract for the radial profiles using a single functional form to capture the average effect.

\begin{figure*}
    \centering
    \includegraphics[trim=1cm 1cm 1cm 0cm,width=\textwidth]{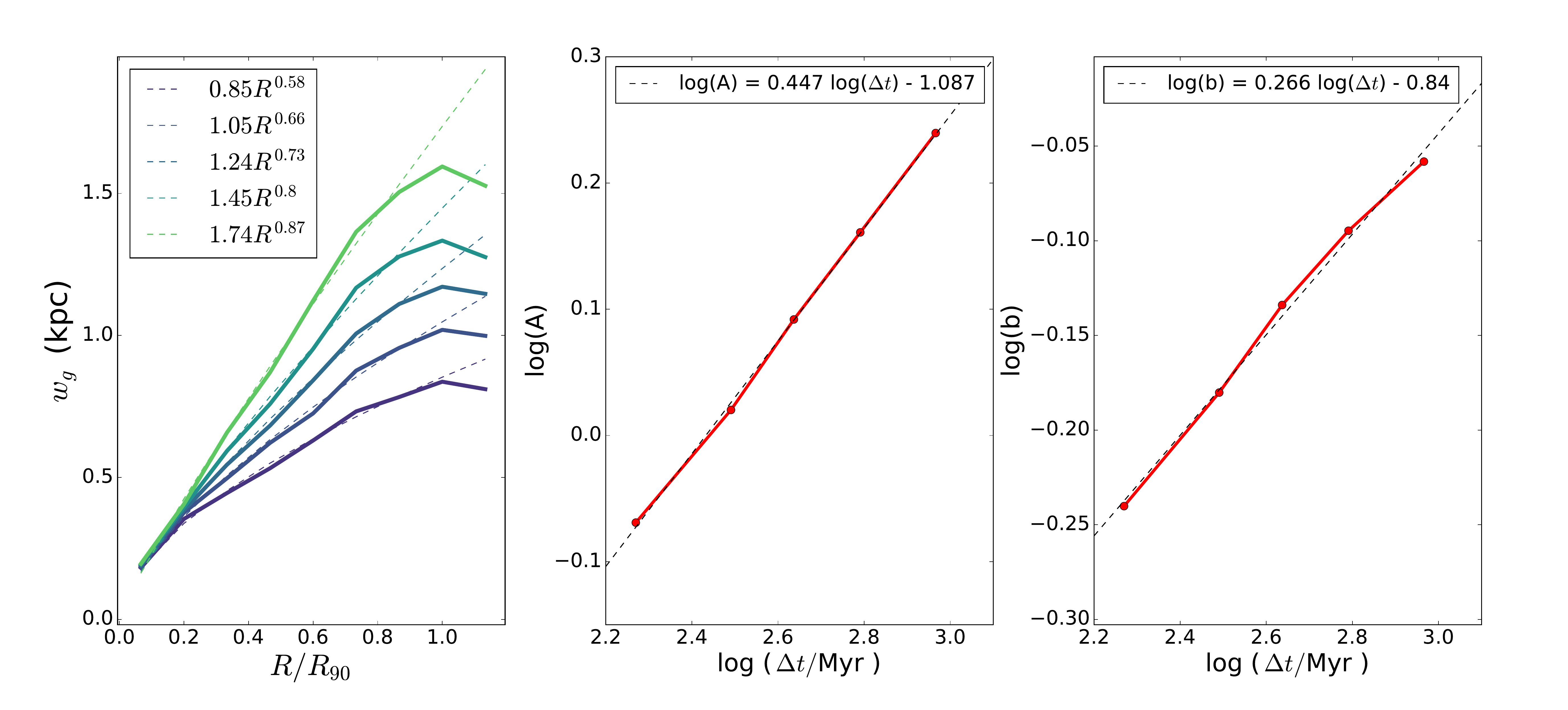}
    \caption{Left panel: Power law fits to the radial profiles of the spread $w_{g}$, referring to the spread of the $\Delta R_g$ histograms, each curve showing a different timestep identical to Figure \protect\ref{rad_prof_snap}. We show the exact values for the fit in the legend. Middle and right panels: We show the logarithmic plots of the coefficient (middle) and exponent (right) of the power law against the time interval $\Delta t$ as well as the best fit line through the data points with its functional form stated in the legend.}
    \label{wg_powerlaws}
\end{figure*}

We fit the radial profiles of $w$ and $w_g$ with power law fits of the form given by Eq.~\ref{pargen} below, with the time interval dependence entering both in the exponent and the normalization of the power law.

\begin{equation}
    w;w_{g} = A(\Delta t) R\sub{sc}^{b(\Delta t)} ,\\
\label{pargen}
\end{equation}

where $A (\Delta{}t) = \Delta{}t^{a}/C_1$ and $b(\Delta{}t) = \Delta{}t^{\beta}/C_2$. In Fig.~\ref{wg_powerlaws} we show fits for the 5 selections of time intervals, and how the exponents and normalization constants vary with increasing time interval. We find that both the coefficient $A$ and the exponents of the power laws in radius vary with the selected time interval $\Delta t$ in a linear fashion when plotted in a logarithmic plot. So each has a power law like dependence on $\Delta t$. 

We note that we have excluded the $n+1$ case from these fits because it significantly deviates from the trend that the rest of the cases follow. We argue that this is because the timespan of only $\sim 60$ Myr is not long enough to robustly measure the diffusion of star particles driven by radial migration, as it is shorter than one rotational period of the stars at most radii. Once we move to time intervals with widths of a few hundred Myr, we find a time evolution that can be expressed accurately with the same power law fit. 

The equations that then express $w$ and $w\sub{g}$ are:

\begin{equation}
    w\sub{g}/\tn{kpc} = \frac{\Delta t^{0.45}}{12.2\,\textrm{Myr}} R\sub{sc}^{(\Delta t^{0.27}\,/\,6.9\,\textrm{Myr})}
\label{parwg}
\end{equation}

\begin{equation}
    w/\tn{kpc} = \frac{\Delta t^{0.18}}{1.2 \, \textrm{Myr}} R\sub{sc}^{(\Delta t^{0.1}\,/\,2.3\,\textrm{Myr})} 
\label{parw}
\end{equation}

where $R\sub{sc} = R/R_{90}$.

As mentioned before we must caution that the exact coefficients and exponents in these equations strictly describe the average of a diverse set of galactic disks and they would be different if we had only considered a particular type of systems such as those with a strong bar. In particular we find that regarding the quantity $w_g$ the best fit exponent $\beta$ in the radial term has the same value for both the subsamples of weakly and strongly barred galaxies in our simulations. However, the exponent $\alpha$ that regulates the timestep dependence in the term $w_g\sim{\Delta t^{\alpha}}$ has a value of 0.38 for the strongly barred subsample and 0.5 for the weakly barred one, these values being on either side of the value of 0.45 for the whole sample. Hence, we can conclude that the presence of the bar does not influence significantly the shape the radial profile of $w_g$ , regulated by the term $b(\Delta t)$, however it has an effect on how the normalization term $A(\Delta t)$ increases with the timestep and in particular leads to slower-than-diffusion behaviour. Regarding the quantity $w$ we recover the same exponent in the dependence $w\sim{\Delta t ^{\alpha}}$ for either subsample as well as for the whole sample.

\begin{figure*}
    \centering
    \includegraphics[trim=1cm 1cm 1cm 0cm,width=\textwidth]{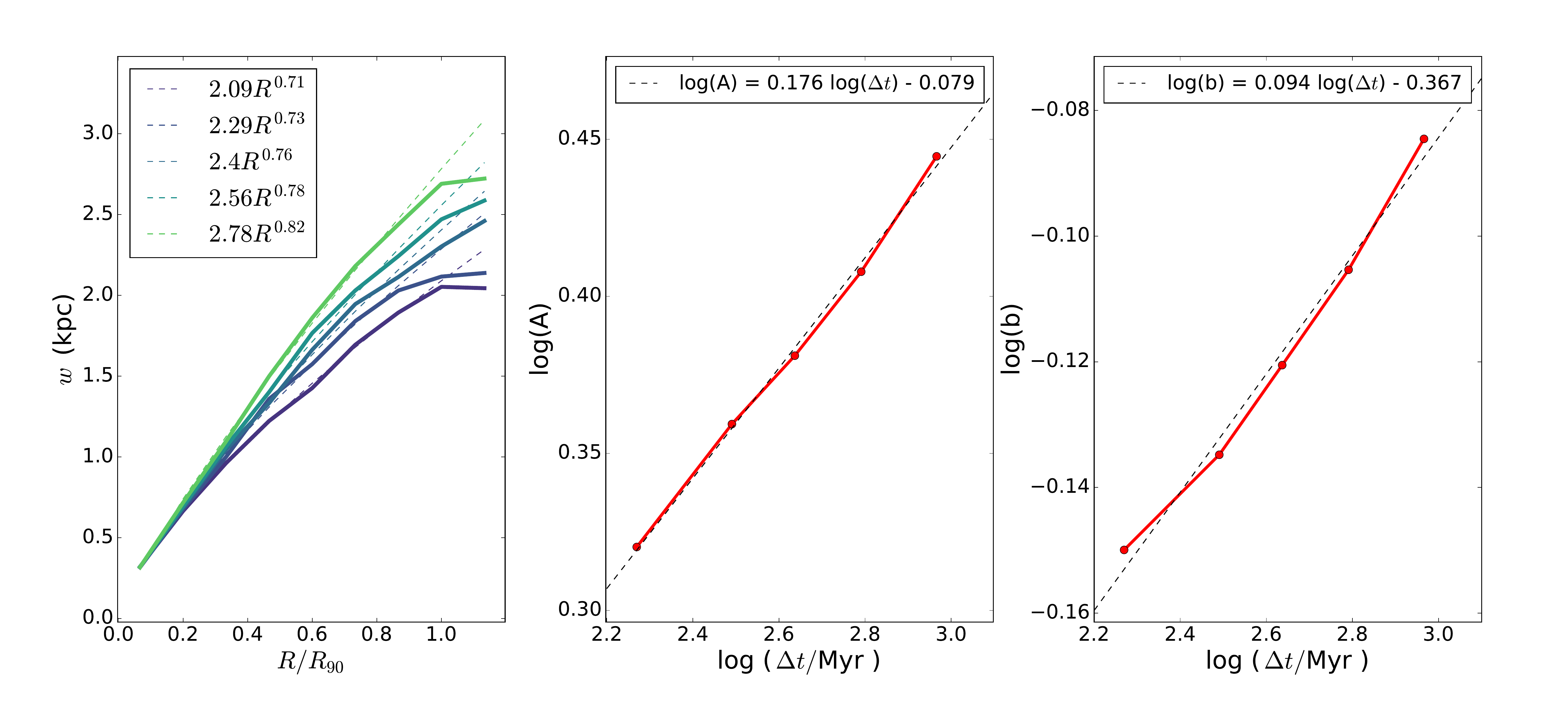}
    \caption{Similar to Fig.~\ref{wg_powerlaws}, but for the quantity $w$ which refers to the spread of $\Delta R$ histograms.}
    \label{w_powerlaws}
\end{figure*}

\section{Discussion and Conclusions}

We have investigated the effect of stellar radial migration in 17 disk galaxies from the Auriga simulations. We consider a narrow sample of disk stellar mass, $10 < \tn{log}(M_{*}/\Msun) < 11$, but with diverse properties in disk kinematics, radii and bar strengths, as well as different evolutionary histories in terms of merger events. 

We have measured the amount of radial migration in our simulations by (1) comparing the radii of disk stars at $z=0$ with their birth radii, and (2) comparing changes in galactocentric radii and orbital guiding centres between pairs of snapshots spanning varying periods of time. The former study allows us to make conclusions about the cumulative effect of migration in the observable quantities of the disk. We note that naively averaging over the difference between birth and present day radius yields values close to zero but in further analysis we see that this is due to the net effect of stars moving both inwards and outwards from their birth positions. In systems with strong bars, we find an excess of ``positive migrators'' (Fig. \ref{rbirth_rfin}), indicating stars that have been `pushed' to the outer disk because of the interaction with the bar. This feature is similar to that shown in \cite{Roskar08b} (although their simulated disk does not develop a bar). \cite{Minchev14} also presents a similar result for the guiding radii of the stars. We have reproduced our Fig. \ref{rbirth_rfin} in terms of guiding radii instead of galactocentric radii but the differences are negligible.

We have also probed into the relation of the age of the stars at redshift zero to the amount that they have migrated using the histograms of $\Delta R = R_{z=0} - R\sub{birth}$ in different age and radius bins, quantifying the migration as the width of these distributions. For some systems (\eg{}those with quieter merger histories), we find that these histograms can be reasonably fit with a Gaussian function. However, for other systems, we find deviations in the wings of the distributions (especially in the inner regions of our simulated discs) which are not conducive to a simple Gaussian fit. Nevertheless, the histograms widen with both increasing stellar age and larger radius for all rings in the studied galaxies. We found average values of the age-$\sigma_{\rm migr}$ dependence that are consistent with the model proposed by \cite{Frankel20}; the predicted range of values of this model lie within the scatter of our simulated relations, albeit with a slightly steeper slope compared to our median relation.

We further identify that there is a radial dependence in the normalization and position of the peak of the distributions with respect to the centre of the initial ring of selection. \cite{Verma21}, using some of the same Auriga halos in their study, find upper limits for the radial migration of 2.21 kpc for stars with age less than 4 Gyr and 3.7 kpc for stars with ages between 4-8 Gyr in the solar neighbourhood. Although we do not use the same age bins we find that for younger stars (<3 Gyr) the mean migration strength varies between 1-2.5 kpc depending on the radial bin. 

In Auriga, the stellar age radial profiles are not significantly affected by migration in a considerable number of the cases. However, we do observe in a subset of the systems a flattening of the age gradients at larger radii.

In terms of the metallicity gradients (Fig. \ref{metallicity_prof}), we observe a dependence similar to \cite{Minchev13} (Fig. 5 of that paper) where the gradients for the younger stellar populations are at most marginally affected by migration but for the older stars there is a more significant flattening in the majority of cases. This age effect is clearly shown in our simulations and we further present possible correlations of the amount of flattening with the stellar mass and the strength of the bar in the disk. However, overall metallicity profiles (\ie{}accounting for stars of all ages) do not appear to evolve strongly with cosmic time in Auriga.

In the second part of our analysis, we focused on pairs of snapshots spaced by a time difference $\Delta t$, so that we can extract information about how the migration process proceeds in given time intervals. We allow the $\Delta t$ to vary between 60 Myr and 1 Gyr. The two measures that we use to describe the histograms, the median shift ($\overline{\Delta r}$ or $\overline{\Delta r_g}$) and the spread ($w$ or $w_g$), can be used in combination to give a description of the migration process. We find that the values for the median shifts are in any case smaller compared to the spread and highly variable from halo to halo, indicating a secondary effect. However, we deem that is more physically motivated to use the quantities $w$ and $w_g$ as main indicators of the migration since this scatter measure contains more information about the radial motions of the stars which are selected in the given ring compared to the median. Still, it must be noted that we find average negative values for the median shift radial profiles, showing a small inwards median motion of the selected stars in the given rings. Although this looks contradictory to the standard picture of stars migrating to and populating the outer parts of disks, it is reconciled by considering that there is always a rather symmetric spread around the median value meaning that we have in all cases a considerable number of stars that have migrated outwards between two snapshots.
  
We find that, when considering the changes of the guiding centres of the stars, there is a migration process that follows a diffusion-like evolution. The exponent of the timestep dependence is not exactly 0.5, as would be expected for pure diffusion, but slightly lower. This `slower' diffusion could be attributed to the fact that we average over our different systems with different structural properties and lifetime evolutions. In individual halos, we do observe exponents that closely match 0.5. As stated before the diffusion exponent is recovered if we consider only the weakly-barred halos in our sample but the strongly barred systems seem to be regulated by a slower-than-diffusion timestep evolution following $w_g\sim\Delta t^{0.38}$. Indeed, most of the halos in Fig. \ref{wg_timefit}, that have exponents closer to 0.5 do not have a strong bar in their centre. The mechanism due to which the presence of a strong bar leads to this behaviour could be explore in a future study. The corresponding time evolution of the changes in the galactocentric radii is flatter and less clear than the respective one for the guiding centres. Although it can still be parametrised, we are cautious about any strong statements on the physical significance of this result because there are several processes that could contribute to changes in the galactocentric radii that would require much deeper analysis to disentangle them cleanly.

The parametrizations we present in Eq. \ref{parwg},\ref{parw} describe how strongly stars migrate out of a ring as a function of the radial location of the ring as well as the timestep that is used to between the initial and final observations of the stellar positions. However, as shown in Fig.~\ref{rad_prof_agebar}, there are secondary dependencies that contribute to scatter about the median relation that we provide. For this figure, we have used a fixed time interval of $n+10$, but the dependence is similar for all other selections of $n+m$ when $3 < m < 15$. In the case of considering changes in guiding centres, we find that our sub-sample of barred systems has consistently higher values of $w\sub{g}$ than the weakly-barred systems, of the order of 30 per cent in the middle parts of the disk. We must caution that the criterion we choose to distinguish between strongly/weakly barred systems (namely, systems with $A_{2}$ above/below 0.3) is at some level arbitrary. However, we have tested that if we instead split the sample into three sub-samples with max(A$_2)$ < 0.2 as weakly barred systems,  0.2 < max(A$_2)$ < 0.4 as intermediate and max(A$_2)$ > 0.4 as strongly barred, we find that the mean radial profiles of $w\sub{g}$ and $\overline{\Delta r\sub{g}}$ for the intermediate sub-sample lie in between the other two. This indicates that our results are consistent with a continuous dependence based on the value of the bar strength. The stellar age adds another source of scatter, with stars in the youngest age bin showing higher values by 20-40 per cent, depending on the radius, compared to those in the oldest age bin. This is similar for both the strongly and weakly barred sub-samples, implying that this source of scatter is independent of the presence of a bar. 

In terms of the overall description of radial migration, we do not give distinct parametrisations that each describe the effects of `churning' and `blurring' as it has been presented, for example, in \cite{Schoenrich09} but we can indirectly associate our computed quantities with these suggested modes of stellar migration. The quantities $w\sub{g}$ and $\overline{\Delta r\sub{g}}$ show the amount of change of the guiding centres, hence the change of orbital angular momentum of the stars, which is predominantly related to the process of `churning', however the $w$ and $\overline{\Delta r}$ are more general and incorporate information about all possible processes that can result in the change of the orbital radius of a stellar particle.

\begin{figure*}
    \centering
    \includegraphics[trim=3cm 1cm 3cm 0cm,width=\textwidth]{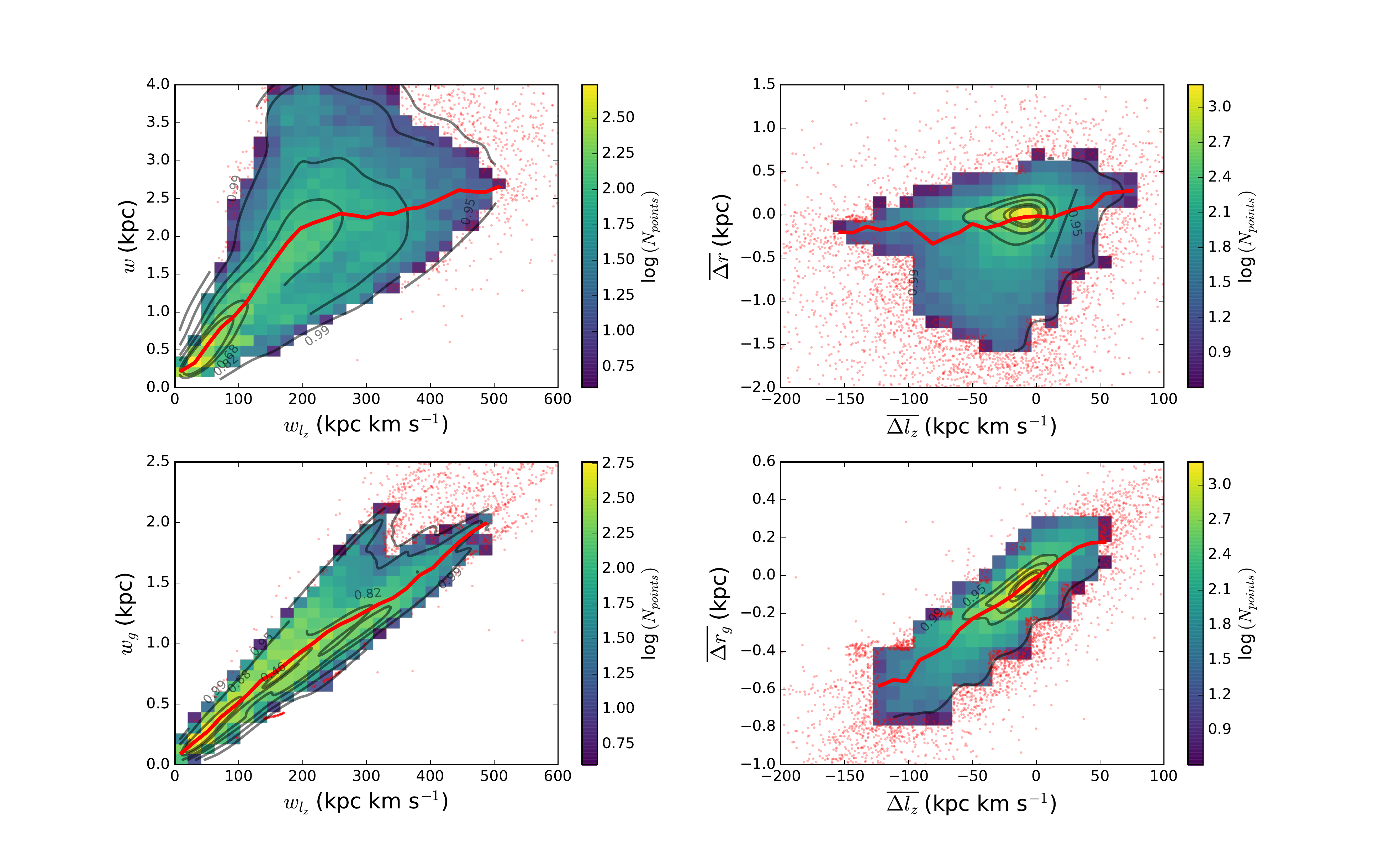}
    \caption{The quantities describing the histograms of $\Delta r$ (top) and $\Delta r_g$ (bottom) plotted against similarly obtained quantities describing the histograms of the change in angular momentum $\Delta l_z$ of the stars. We overplot the median lines in red and the density contours in the density colourmaps while the red data points lie outside the 99th percentile contour. We find very strong correlations between the guiding centre changes and angular momentum changes. This is not true for the galactocentric radii changes, and in particular there is hardly any correlation between the median shifts $\overline{\Delta r}$ and $\overline{\Delta l_z}$.}
    \label{angularmom}
\end{figure*}

Indeed, we find that if we create similar histograms in terms of the changes in angular momentum $\Delta l_{z}$ and extract the related quantities $w_{l_{z}}$ and $\overline{\Delta l_z}$, we see that there is a very tight correlation between the corresponding quantities for the guiding centres, as shown in Fig.~\ref{angularmom} This is expected, as by construction the change of the guiding centre is due to a change in the angular momentum of the star. A correlation between $w_{l_{z}}$ and $w$ is present, although looser than the one with respect to $w\sub{g}$. This is an indication that the quantity $w$ encapsulates the information from $w\sub{g}$, as well as some additional scatter that can be associated with the `blurring' process, which describes the changes in the epicyclic amplitudes on top of the effects that change the angular momentum of the star. On the other hand, there is no correlation at all between $\overline{\Delta l_z}$ and $\overline{\Delta r}$. Whereas there is a median change in the galactocentric radius of the stars selected in the ring, this does not result exclusively from a change in the angular momentum, and this is why a correlation does not arise.

Finally, it must be noted that the parametrisations that we extract are calibrated in our available sample, therefore they are describing the migration process at systems that are comparable in size to our Milky-Way but may not be readily available to disc galaxies of different mass range. 

All in all, in this study we find that:
\begin{itemize}
    \item The average change in the radius for a stellar particle over its lifetime, $\langle{} \Delta R \rangle{}$, is close to zero for all our systems. However, this is due to the fact that stars experience both negative and positive migration in comparable amounts which cancel out to a close-to-zero average. The typical star experiences a migration of the order of 2 kpc.
    \item At $z=0$, we find that older stars (9-12 Gyr) have experienced up to twice the amount of migration compared to newly formed stars (0-3 Gyr). We find a clear age dependence in the migration strength as well as a radial dependence. Stars that have been born at larger radii show broader distributions of $z=0$ radii, regardless of their $z=0$ age.
    \item Stellar migration has a varied effect on the age profiles of the disks at $z=0$, depending on the particular halo, but it does not affect the scatter around the mean stellar age at a given radius.
    \item There is no imprint of migration on the total metallicity profiles or the profiles for young stars (< 3 Gyr), but we find significant flattening of the profile gradients in many systems for the older stellar populations. The extent of the flattening is correlated with the presence of a bar in the disk.
    \item We create distributions of the change in the galactocentric radius $\Delta r$ and the change in guiding centre $\Delta r_g$ between two simulation snapshots for the stars within different annuli and quantify them using a measure of their spread $w$ and $w_g$, and a measure for their median $\overline{\Delta r}$ and $\overline{\Delta r_g}$, respectively. This shows that $w$ and $w_g$ have a power law radial dependence, increasing in the outer regions of the disks as well as a dependence in the time interval between the two snapshots. We present parametrisations that describe these effects. $w_g$ appears to approach a diffusion process at the outermost rings but there is significant halo-to-halo variability.
    \item $\overline{\Delta r_g}$ correlates exactly with changes in the orbital angular momentum $\overline{\Delta l_z}$ of the stars, as expected in the `churning' process. $\overline{\Delta r_g}$ is uncorrelated to $\overline{\Delta l_z}$, being a more random measure that includes the additional effect of `blurring'.
    \item Combining the findings presented in Figs.~\ref{metallicity_slopes},\ref{drg_rg}, and \ref{rad_prof_agebar} we argue that in our sample the systems with a stronger bar are associated with a stronger migration of the stellar particles. This is manifested both in terms of larger values in the changes of guiding radii, $\Delta r_g$ as well as producing shallower slopes in the metallicity profiles for older stellar populations.

\end{itemize}

We note that our results are subject to the limitations of the Auriga galaxy formation model. While dynamical interactions of the stellar particles with the bar are accurately captured, smaller scale effects that could result in stellar migration, such as scattering with molecular clouds, cannot be accounted from the modelling of the ISM.
In future studies we aim to introduce and test the parameterisations for stellar migration in the latest version of the \textsc{L-Galaxies} semi-analytic model of galaxy formation \citep{Henriques+20,Yates+21a}. Here, we present a basic form of such possible parameterisations, focusing on the radial dependence of the migration strength which can be implemented directly into the radial ring model of \textsc{L-Galaxies}. This will allow stars to migrate from ring to ring based on the radial position in each snapshot. As we mention above, the secondary effects of stellar age or bar strength could be also implemented as a scatter around the median radial dependence. Furthermore, it would be of interest to extend this study to disks of smaller masses in order to compare our findings. 

\section*{Acknowledgements}
We thank the anonymous referee for the very constructive report which helped in improving this manuscript. 
RG acknowledges financial support from the Spanish Ministry of Science and Innovation (MICINN) through the Spanish State Research Agency, under the Severo Ochoa Program 2020-2023 (CEX2019-000920-S).
Part of this research was carried out on the High Performance Computing resources at the Max Planck Comput-ing and Data Facility (MPCDF) in Garching operated by the MaxPlanck Society (MPG).

\section*{Data Availability}

The data underlying this article will be shared on reasonable request to the corresponding author.



\bibliographystyle{mnras}
\bibliography{stellar_migration_Auriga} 




\appendix

\section{Additional figures}

\begin{figure*}
    \centering
    \includegraphics[scale=0.5]{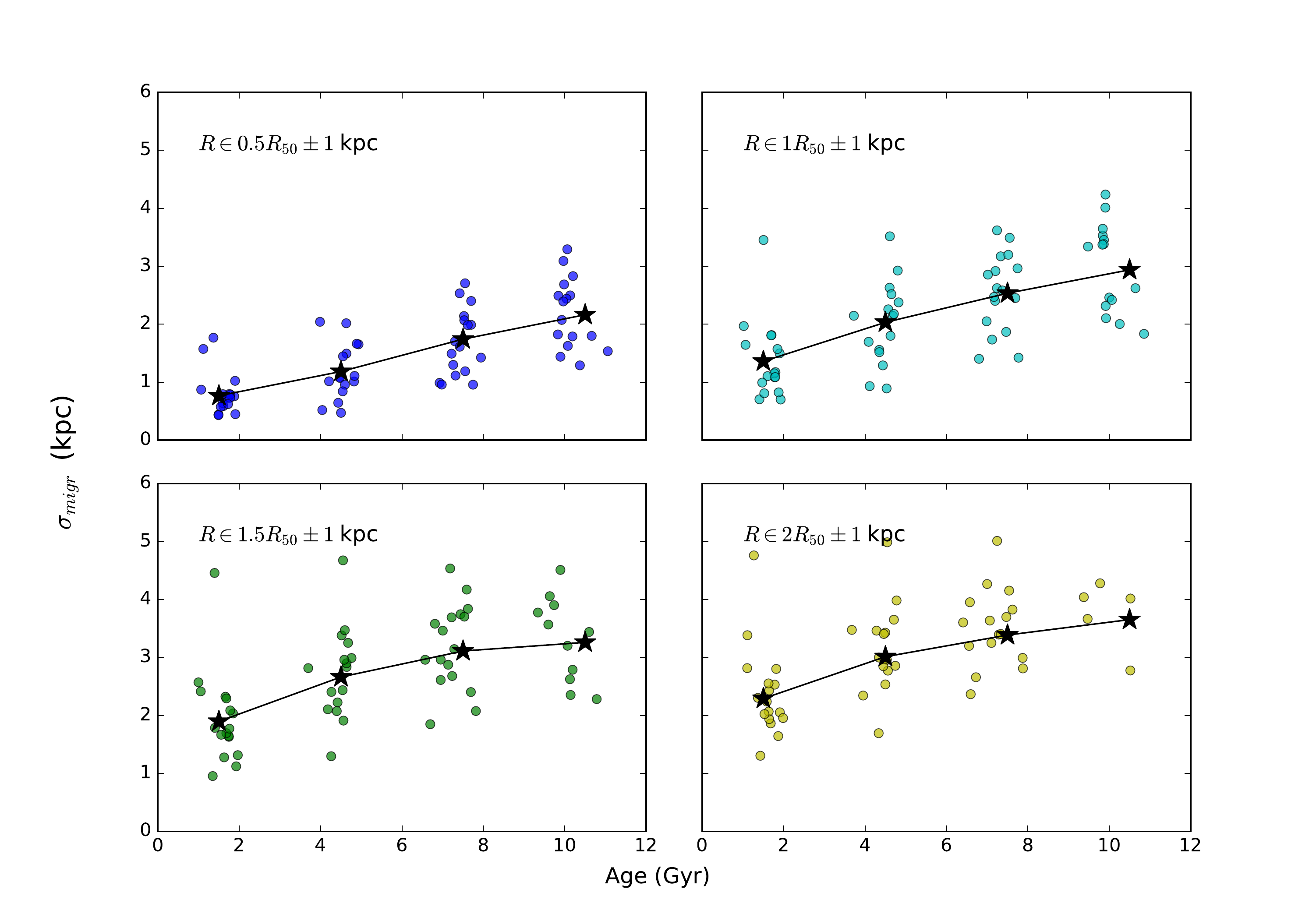}
    \caption{The information used in Fig.~\ref{migr_age_medians} presented in separate axis, showing  the data points that are used to calculate the median curves and the errors.}
    \label{migr_age}
\end{figure*}

In Fig. ~\ref{migr_age} we plot the migration strength $\sigma_{migr}$ versus the age of the stars at $z=0$. We use four radial bins, each shown in the four panels and we also split the stars based on their age at $z=0$ in four further bins. In this figure each datapoint is drawn from a separate halo and represents the width of histograms, such as those presented in \ref{halo6_hist}. The same information is conveyed in Fig. \ref{migr_age_medians} in a more concise presentation, showing the scatter of the data points with error bars around the median curves. 

\begin{figure}
    \centering
    \includegraphics[scale=0.4]{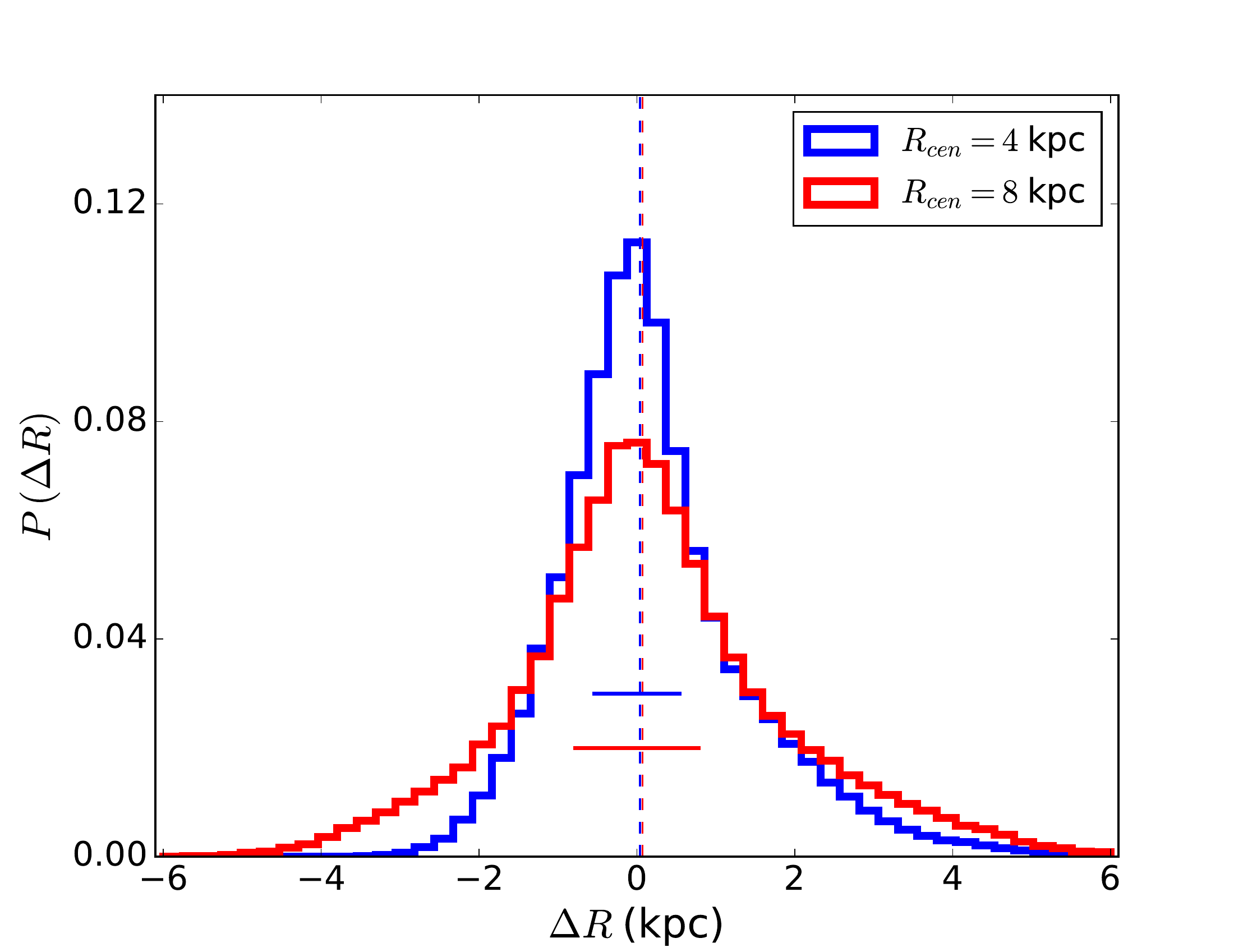}
    \caption{Example distributions of the change of the galactocentric radii for the stars selected at two given rings centred at 4 (blue) and 8 (red) kpc for `halo\_6'. From these histograms we compute the median shift ($\overline{\Delta r}$) shown with the dashed line, and the 16-84 percentile range ($w$) shown by the horizontal lines. We observe that for the outer ring the distribution is more broadened. In both rings the shift of the median from zero is very small, and in these particular examples is slightly positive.}
    \label{nmhist}
\end{figure}

In Fig.~\ref{nmhist} we give examples of how the histograms in $\Delta R$ look like. From such histograms we extract the median ($\overline{\Delta r}$) and the width ($w$) which we use to describe stellar migration for the stars in the given ring. Histograms in terms of $\Delta R_g$ look exactly similar.

\begin{figure*}
    \centering
    \includegraphics[trim=3cm 1cm 3cm 0cm,width=\textwidth]{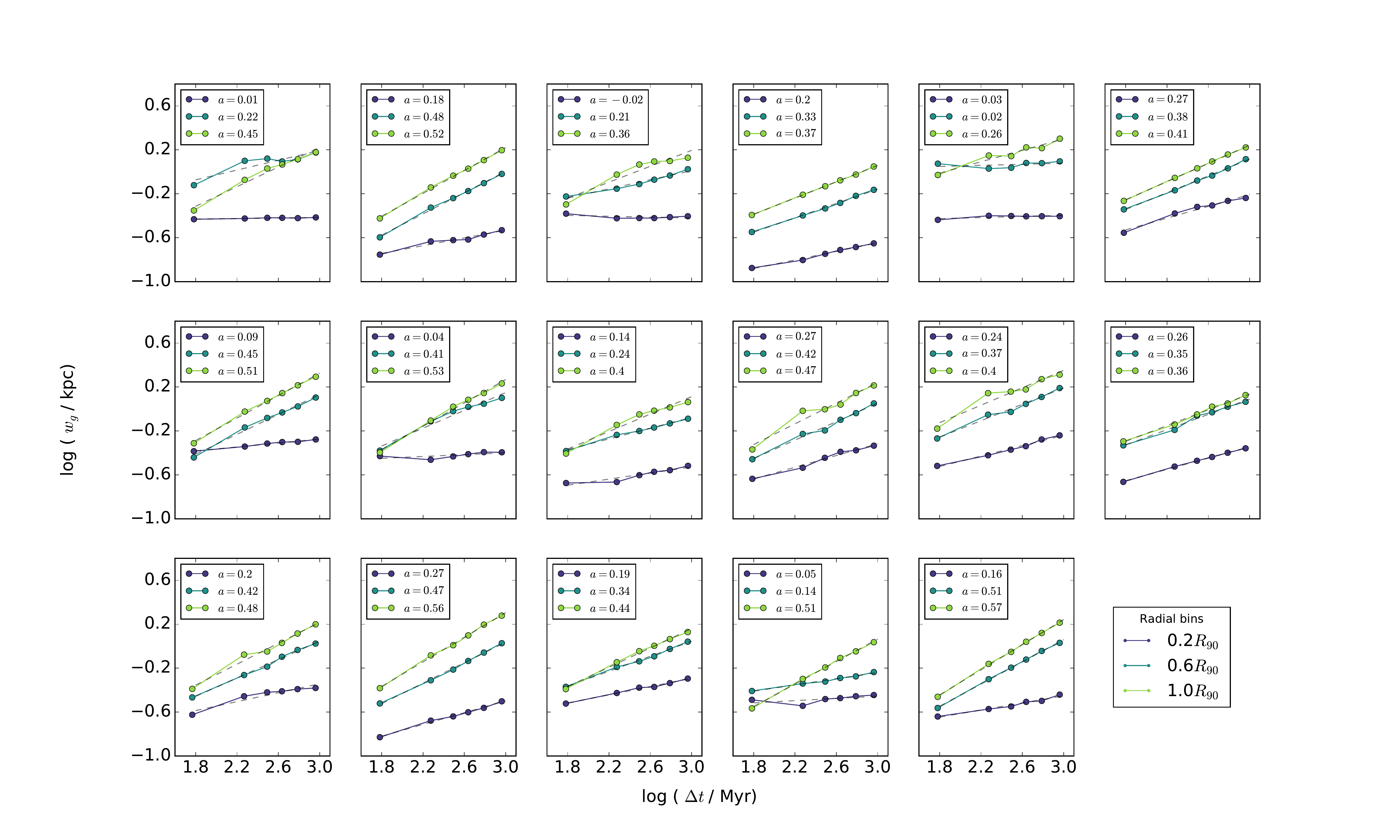}
    \caption{Logarithmic plots of the spread $w_{g}$ against the time interval $\Delta t$ for individual halos. The different curves are for different normalised radii within the disks. The slope of the best fit line is quoted in the legend in each panel. We find a variety of different values for the slopes, ranging between 0.3-0.6 in most galaxies for the two outermost rings (cyan and green).}
    \label{wg_timefit}
\end{figure*}

In Fig.~\ref{wg_timefit} we show the time interval dependence of the quantity $w_g$, similar to Fig.~\ref{wg_timefit_all}, for each individual galaxy in three different radial bins. We notice that in some disk, in the two outermost rings the value of the slope is near or around the diffusion value of 0.5.

\bsp	
\label{lastpage}
\end{document}